\pdfoutput=1

\documentclass[11pt,letterpaper]{article}

\usepackage[T1]{fontenc}
\usepackage[utf8]{inputenc}
\usepackage{lmodern}
\usepackage[margin=1in]{geometry}
\usepackage{amsmath,amssymb}
\usepackage{graphicx}
\usepackage{booktabs}
\usepackage{array}
\usepackage{multirow}
\usepackage{float}
\usepackage{placeins}
\usepackage[font=small,labelfont=bf,labelsep=colon]{caption}
\usepackage{enumitem}
\usepackage{xcolor}
\usepackage{tikz}
\usetikzlibrary{positioning,arrows.meta,calc,fit}
\usepackage{microtype}
\usepackage[hidelinks]{hyperref}
\hypersetup{
  pdftitle={Domain-repetition patterns in protein domain trigrams are uniform across chromosomes and conserved among multicellular eukaryotes},
  pdfauthor={Yanying Wu},
  pdfsubject={Comparative genomics; protein domain architecture},
  pdfkeywords={protein domains; Pfam; domain architecture; trigram; n-gram; Jensen-Shannon distance; chromosomes; comparative proteomics}
}

\graphicspath{{figures/}}

\definecolor{tAAA}{HTML}{2A78D6}
\definecolor{tAAB}{HTML}{EB6834}
\definecolor{tABB}{HTML}{1BAF7A}
\definecolor{tABA}{HTML}{EDA100}
\definecolor{tABC}{HTML}{E87BA4}
\newcommand{\ty}[1]{\textsf{#1}}
\newcommand{\pf}[1]{\texttt{#1}}
\newcommand{\org}[1]{\textit{#1}}
\newcommand{\fname}[1]{{\scriptsize\ttfamily\color{blue!55!black}#1}}

\setlist{itemsep=2pt,topsep=4pt}

\title{\textbf{Domain-repetition patterns in protein domain trigrams are uniform across chromosomes and conserved among multicellular eukaryotes}}
\author{Yanying Wu\\[4pt]
  {\normalsize\itshape Centre for Neural Circuits \& Behaviour, University of Oxford, Oxford, United Kingdom}\\
  {\normalsize\texttt{yanying.wu@cncb.ox.ac.uk}}}
\date{12 November 2026}

\begin{document}
\maketitle

\begin{abstract}
\noindent
Proteins are built from domains, and the order of domains along a protein can be read like a sentence in which each
domain is a word. We take every run of three consecutive Pfam domains (a domain trigram) and classify it by its pattern
of repetition: all three the same (\ty{AAA}), the first two the same (\ty{AAB}), the last two the same (\ty{ABB}), the
outer two the same (\ty{ABA}) or all different (\ty{ABC}). We then ask whether the proportions of these five patterns
are conserved across the chromosomes of a genome and across genomes.

We built a reproducible six-step workflow from protein sequence to trigram proportions and applied it to the current
UniProt reference proteomes of 16 organisms, from \org{Escherichia coli} to human, with Pfam~39.0 domain annotation. In
the fruit fly \org{Drosophila melanogaster}, 3{,}107 distinct trigrams split into 63\% \ty{ABC}, 13\% each of \ty{AAB}
and \ty{ABB}, and 5\% each of \ty{AAA} and \ty{ABA}. Every chromosome arm closely matches this pie: each lies within the
range expected for a random set of genes of the same size. The human pie has less \ty{ABC} (55\%) and more \ty{AAB} and
\ty{ABB} (18\% each), and 23 of its 25 chromosomes and other components fall within the random-gene range. Pies of trigram
occurrences vary more, but most of that variation traces to a few giant repeat proteins (\textit{dumpy} gives 34\% of the
occurrences on fly arm 2L).

Across 199 chromosomes and other proteome components of 14 organisms, tested with a gene-level permutation, 4 (2\%) differ
from their own proteome in distinct trigrams and 10 (5\%) in occurrences after correction, chiefly human and macaque
chromosome~19, whose KRAB zinc-finger clusters make it rich in \ty{AAA} repeats. Among the genomes other than fly and
human, the chromosomes of \org{S.~pombe}, \org{Arabidopsis} and maize lie closest to their own proteome, and those of
\org{S.~cerevisiae}, \org{E.~coli} O157:H7 and \org{C.~elegans} furthest; none of the latter three has a component that
differs significantly, and the raw distance largely reflects how few trigrams a component holds. Across organisms, the
distinct-trigram pie is closely conserved among multicellular eukaryotes (median pairwise distance 0.036), with \ty{ABC} at
55--67\% and \ty{AAB} and \ty{ABB} always within 2 percentage points of each other; it shifts towards \ty{ABC} in yeasts
(75--77\%) and bacteria (82--86\%).

The proportion of domain-repetition patterns is therefore conserved across the chromosomes of a genome and shared by plants
and animals, while repeat-rich giant proteins and tandem gene clusters account for most of the local deviations.
\end{abstract}

\section{Introduction}

Most eukaryotic proteins contain more than one domain\footnote{\textit{Domain}: an evolutionarily conserved,
usually independently folding unit of protein sequence and structure. Here domains are Pfam families, each
defined by a profile hidden Markov model \cite{paysanlafosse2025pfam}.}, and the order in which the domains occur, the
\emph{domain architecture}, is a primary determinant of what a protein does \cite{vogel2004multidomain,chothia2003repertoire}.
Architectures evolve by duplication, fusion, fission and shuffling of domains \cite{bjorklund2005rearrangements,moore2008arrangements,kummerfeld2005fusion},
and by the internal duplication of a single domain into tandem repeats \cite{andrade2001repeats,bjorklund2006repeats,marcotte1999census}.
Their distribution is far from random: a few domains combine with many partners while most combine with few
\cite{apic2001combinations}, and the proportion of multidomain proteins differs between the three kingdoms of
life \cite{ekman2005multidomain}. Genes are not placed at random along chromosomes either \cite{hurst2004order}, and
chromosomes differ widely in length, gene density and history. It is therefore not obvious that different chromosomes, or
different genomes, should use domain architectures in the same way.

Because a domain architecture is a string of symbols, methods from natural language processing apply naturally to
it. Statistical language models of domain order improve domain detection \cite{coin2003language}, and an analysis of
domain bigrams and trigrams across thousands of genomes found a quasi-universal ``grammar'' of domain architectures, whose
information content changes little between the major lineages \cite{yu2019grammar}. In the simplest $n$-gram
models\footnote{An \textit{$n$-gram} is a run of $n$ consecutive symbols; a \textit{trigram} is a run of three
\cite{jurafsky2009speech}.} a text is summarised by the frequencies of its short runs of words \cite{jurafsky2009speech}, and
the idea that speakers spread information evenly along an utterance \cite{levy2007speakers,jaeger2010redundancy} motivated
us to ask whether such a summary stays constant along a genome.

Here we use the shortest run that can show how a protein repeats and alternates its domains: the trigram of three
consecutive Pfam domains. Three positions are enough to distinguish five patterns of repetition (Fig.~\ref{fig:workflow}b):
a homogeneous repeat (\ty{AAA}), the end of a repeat run (\ty{AAB}), the start of one (\ty{ABB}), a domain sandwiched by
another (\ty{ABA}), and three different domains (\ty{ABC}). The proportions of the five patterns, drawn as a pie,
summarise how a set of proteins mixes repetition and novelty. We ask whether that summary is the same on every
chromosome of a genome, and how much it varies between genomes.

We describe a documented workflow and a statistical test of chromosome-to-proteome similarity that accounts for the
clustering of trigrams within genes (\S\ref{sec:methods}). We then report the results for the fruit fly (\S\ref{sec:fly})
and human (\S\ref{sec:human}), summarise all 16 organisms (\S\ref{sec:summary}), and examine the three genomes, other than
fly and human, whose chromosomes follow their whole-proteome pie most closely (\S\ref{sec:most}) and least closely
(\S\ref{sec:least}). Finally, we test how robust the results are to database updates (\S\ref{sec:2020}).

\section{Data and methods}\label{sec:methods}

\subsection{From protein sequence to trigrams}

\begin{figure}[!tbp]
\centering
\begin{tikzpicture}[
  font=\small,
  stepbox/.style={draw=black!70, rounded corners=2pt, fill=black!4, align=center, text width=6.6cm, minimum height=0.75cm, inner sep=3pt},
  lbl/.style={font=\footnotesize, text=black!75, align=left, anchor=west},
  arr/.style={-{Stealth[length=2.2mm]}, thick, black!70}
]
\node[font=\bfseries] at (-0.3,0.95) {(a)};
\node[stepbox] (s1) at (3.5,0) {Protein sequences of each chromosome\\[-1pt]\fname{<sp>\_<chr>\_translations.fasta}};
\node[stepbox, below=0.45cm of s1] (s2) {Pfam matches of each protein\\[-1pt]\fname{<sp>\_<chr>\_InterPro\_pfam.csv}};
\node[stepbox, below=0.45cm of s2] (s3) {Ordered Pfam domain list of each protein\\[-1pt]\fname{<sp>\_<chr>\_pfam.dat}};
\node[stepbox, below=0.45cm of s3] (s4) {Trigrams and their proteins\\[-1pt]\fname{<sp>\_<chr>\_trigram.dat}};
\node[stepbox, below=0.45cm of s4] (s5) {Trigrams sorted by number of occurrences\\[-1pt]\fname{<sp>\_<chr>\_sorted\_trigrams.dat}};
\node[stepbox, below=0.45cm of s5] (s6) {Trigram types and their proportions\\[-1pt]\fname{<sp>\_trigram\_summary.csv}};
\foreach \a/\b in {s1/s2,s2/s3,s3/s4,s4/s5,s5/s6} \draw[arr] (\a) -- (\b);
\node[fill=black!70, circle, inner sep=2pt, above=0.35cm of s1] (start) {};
\draw[arr] (start) -- (s1);
\node[lbl] at ($(start.east)+(0.2,0)$) {Step 1: UniProt REST API; split proteome by component};
\node[lbl] at ($(s1.south)!0.5!(s2.north)+(0.2,0)$) {Step 2: Pfam 39.0 matches (InterPro API or Pfam release file)};
\node[lbl] at ($(s2.south)!0.5!(s3.north)+(0.2,0)$) {Step 3: \texttt{parse\_pfam()} --- order by start, remove identical sequences};
\node[lbl] at ($(s3.south)!0.5!(s4.north)+(0.2,0)$) {Step 4: \texttt{collect\_trigrams()} --- sliding window of three domains};
\node[lbl] at ($(s4.south)!0.5!(s5.north)+(0.2,0)$) {Step 5: \texttt{sort\_trigrams()} --- count and name};
\node[lbl] at ($(s5.south)!0.5!(s6.north)+(0.2,0)$) {Step 6: \texttt{categorize\_trigrams()} --- \ty{AAA}/\ty{AAB}/\ty{ABB}/\ty{ABA}/\ty{ABC}};
\node[lbl] at ($(s6.east)+(0.2,0)$) {then: pies; similarity to the whole (\S\ref{sec:sim})};

\begin{scope}[shift={($(s6.south)+(-3.5,-1.1)$)}]
\node[font=\bfseries] at (-0.3,0.45) {(b)};
\node[anchor=west, font=\footnotesize] at (0.0,0.45) {Domain list of a myosin VIIa (\texttt{Q9V3Z6}) and its six trigram occurrences};
\def\doms{{"PF00063","PF00612","PF00612","PF00612","PF00612","PF00784","PF00784","PF00373"}}
\def\dcol{{"black!25","blue!20","blue!20","blue!20","blue!20","orange!30","orange!30","green!25"}}
\foreach \i in {0,...,7} {
  \pgfmathsetmacro{\c}{\dcol[\i]}
  \node[draw=black!60, fill=\c, minimum width=1.12cm, minimum height=0.45cm, inner sep=0pt, font=\scriptsize\ttfamily]
    (d\i) at (0.6+\i*1.2, -0.2) {\pgfmathparse{\doms[\i]}\pgfmathresult};
}
\foreach \i/\t/\col in {0/ABB/tABB,1/AAA/tAAA,2/AAA/tAAA,3/AAB/tAAB,4/ABB/tABB,5/AAB/tAAB} {
  \pgfmathtruncatemacro{\j}{\i+2}
  \draw[\col, line width=1.8pt] ($(d\i.south west)+(0.08,-0.22-\i*0.3)$) -- ($(d\j.south east)+(-0.08,-0.22-\i*0.3)$)
     node[right, font=\scriptsize\sffamily, text=\col] {\t};
}
\node[anchor=north west, font=\scriptsize, text=black!70, align=left] at (0.0,-2.35)
  {\pf{PF00063} myosin head;\ \pf{PF00612} IQ calmodulin-binding motif;\ \pf{PF00784} MyTH4;\ \pf{PF00373} FERM central};
\node[anchor=north west, font=\footnotesize, align=left] at (10.9,0.1)
  {\textcolor{tAAA}{\ty{AAA}} all three identical\\
   \textcolor{tAAB}{\ty{AAB}} first two identical\\
   \textcolor{tABB}{\ty{ABB}} last two identical\\
   \textcolor{tABA}{\ty{ABA}} first and last identical\\
   \textcolor{tABC}{\ty{ABC}} all three different};
\end{scope}
\end{tikzpicture}
\caption{The workflow. (a) The six steps from protein sequences to trigram-type proportions, with the file written
by each step (\texttt{<sp>}: species prefix, \texttt{<chr>}: proteome component) and the function that writes it.
Steps 1--2 are in \texttt{fruitfly\_01\_download\_data} (fly) or the first part of each \texttt{<species>\_trigram}
notebook; Steps 3--6 in \texttt{fruitfly\_02\_trigram\_analysis} or the second part. (b) Worked example: the eight Pfam
domains of a myosin VIIa (domain list as in the original workflow) give six overlapping trigram occurrences, coloured
by type. Every run of identical domains that is flanked on both sides produces one \ty{ABB} at its start and one
\ty{AAB} at its end.}
\label{fig:workflow}
\end{figure}
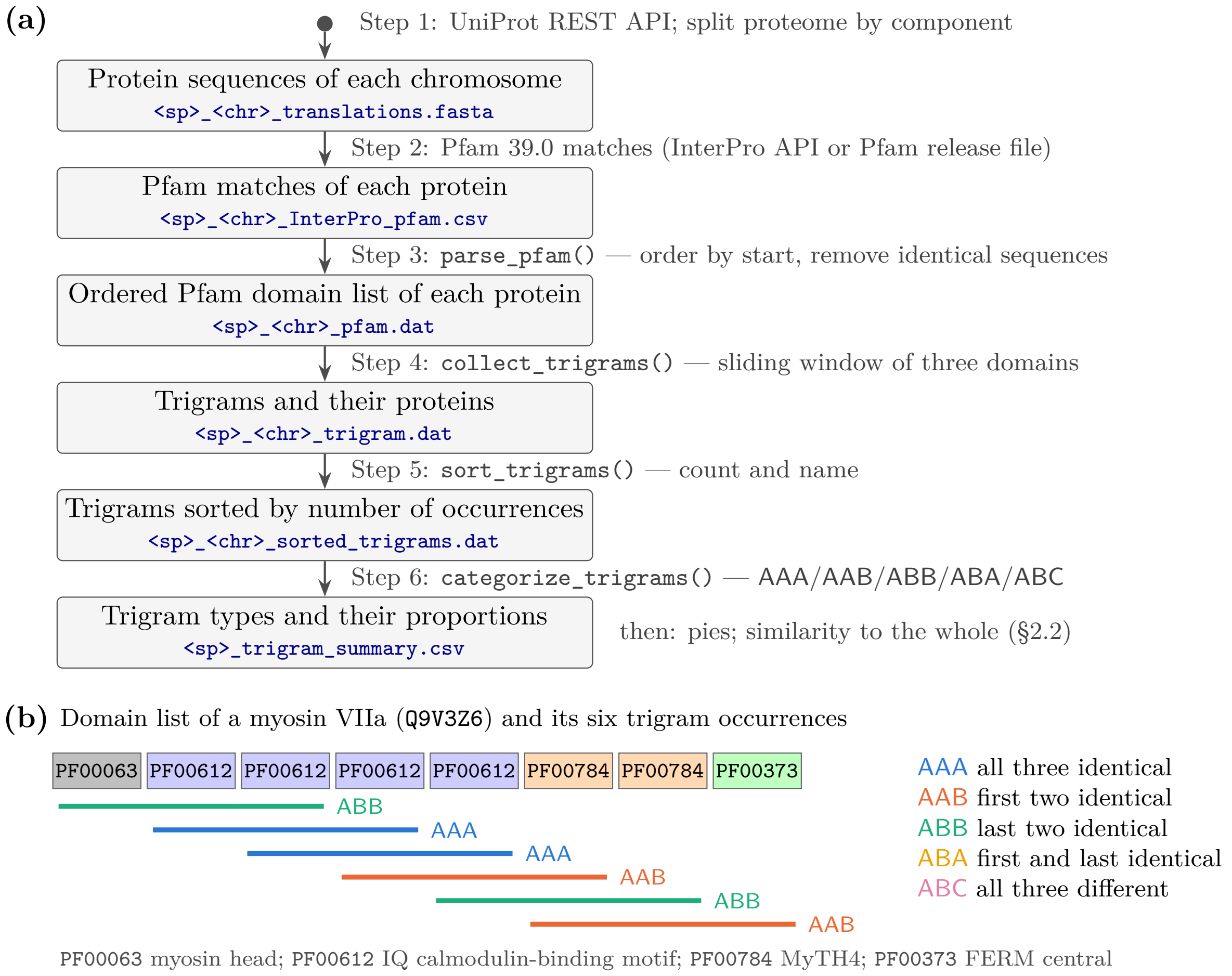

The workflow has six steps (Fig.~\ref{fig:workflow}a), each implemented as a Python function and written to its own
folder, so that every intermediate file can be inspected.

\paragraph{Step 1: protein sequences per chromosome.} For each organism, the whole UniProt reference proteome
\cite{uniprot2025} (release 2026\_04) was downloaded through the UniProt REST API together with the \emph{proteome
component} of each entry --- the chromosome, organelle genome, plasmid or unplaced scaffold that encodes it --- and the
FASTA file was split by component. Two labelling errors of the current release were corrected: about 170 fly
chromosome~4 proteins labelled \emph{Mitochondrion} were moved back to chromosome~4 unless their gene is a
mitochondrial \texttt{mt:} gene (giving 200 chromosome~4 and 13 mitochondrial proteins), and 65
\org{Arabidopsis} mitochondrial proteins labelled \emph{Chromosome~2} were moved to the mitochondrion using the
companion organelle proteome.

\paragraph{Step 2: Pfam domains.} Pfam~39.0 \cite{paysanlafosse2025pfam} domain matches came from one of two routes.
For the fly and six smaller proteomes (two \org{E.~coli} strains, both yeasts, \org{Tetrahymena} and \org{C.~elegans}),
InterPro~111.0 \cite{blum2025interpro} was queried for every Pfam entry occurring in the proteome; InterPro runs
InterProScan \cite{jones2014interproscan} on all of UniProtKB, so this is equivalent to running
\texttt{interproscan.sh -appl Pfam} locally, which remains an option in the notebooks. A domain matched in several
fragments was recorded from its first start to its last end. For the nine large plant and vertebrate proteomes,
the InterPro API timed out (HTTP~408) on every request, so matches were read from the Pfam~39.0 release file
\texttt{Pfam-A.regions.tsv.gz}, joined to UniProt entries by sequence MD5 checksum and placed by their alignment
coordinates. Entries whose sequence is newer than the Pfam release (dated after 30 September 2026) were never scanned
by Pfam and were excluded (Table~\ref{tab:species}); this matters only for mouse (174{,}060 entries, mostly isoform-like
entries added in 2026) and human (116{,}508). On the six non-fly API organisms the two routes give identical domain
coordinates for 94--99.7\% of locations (\S\ref{sec:caveats}).

\paragraph{Step 3: domain list per protein.} The Pfam domains of each protein were ordered by start position. Identical
sequences (same MD5) were kept once, preferring the reviewed Swiss-Prot entry; this removes 892 of 18{,}373 Pfam-bearing
fly proteins, mostly TrEMBL duplicates of Swiss-Prot entries.

\paragraph{Step 4: trigrams.} A window of three consecutive domains was slid along every protein with at least three
domains. A protein with $k$ domains contributes $k-2$ trigram \emph{occurrences}; a trigram that occurs twice in a protein
counts twice.

\paragraph{Step 5: sorting.} Trigrams were sorted by their number of occurrences and annotated with domain names.

\paragraph{Step 6: trigram types.} Each trigram $(d_1,d_2,d_3)$ was assigned to one of five types: \ty{AAA} if
$d_1=d_2=d_3$; otherwise \ty{AAB} if $d_1=d_2$, \ty{ABB} if $d_2=d_3$, \ty{ABA} if $d_1=d_3$, and \ty{ABC} if all three
differ. Proportions were computed in two ways: over \emph{distinct trigrams}, each different trigram counted once,
and over \emph{trigram occurrences}, every occurrence counted. The first describes the vocabulary of domain
combinations; the second weights each combination by how often it is used, so that repeat-rich proteins weigh more.

\subsection{Similarity of a chromosome to its proteome}\label{sec:sim}

For each component $c$ (chromosome, arm, organelle, plasmid or unplaced set) and for the whole proteome, the
trigram types give a five-part composition, $p_c$ and $p$. Their difference was measured by the Jensen--Shannon
distance\footnote{The square root of the Jensen--Shannon divergence, which is a metric \cite{endres2003metric}; with
base-2 logarithms it runs from 0 (identical compositions) to 1 (no type in common).} \cite{lin1991divergence}
\begin{equation}
  \mathrm{JS}(p_c,p) = \sqrt{\tfrac12 \mathrm{KL}(p_c\,\|\,m) + \tfrac12 \mathrm{KL}(p\,\|\,m)},\qquad m=\tfrac12(p_c+p),
  \label{eq:js}
\end{equation}
where KL is the Kullback--Leibler divergence in bits; we also report $1-\mathrm{JS}$ as a similarity and the type with the
largest difference in percentage points (pp).

Any finite subset of a proteome gives a pie that differs a little from the whole, so the observed distance must be
compared with that of a random subset of the same size. Trigrams are not independent: one repeat protein can contribute
more than a hundred occurrences, and a gene may be represented by tens of isoforms in the proteome (three isoforms of
\textit{dumpy} alone give 333 occurrences on fly arm 2L). A test on trigram counts, or a permutation of proteins, would
treat these as independent and call small, gene-driven differences highly significant. We therefore permuted
\emph{genes}: component labels were shuffled across genes, with all isoforms of a gene moving together, 5{,}000 times
(seed 0), and the distance was recomputed each time. Genes were taken from the \texttt{GN=} field of the UniProt header
(the accession when there is none; 0.2\% of human and 11\% of maize proteins). The permutation $p$-value is
$(1+\#\{\mathrm{JS}_{\mathrm{perm}}\ge \mathrm{JS}_{\mathrm{obs}}\})/(1+5000)$ \cite{phipson2010permutation}, adjusted by
the Benjamini--Hochberg procedure \cite{benjamini1995fdr} over the components of one organism and one measure. We also
report the 95th percentile of the permutation distances (the ``random-gene range''). Components with fewer than ten
genes are kept in the whole-proteome pool but not tested.

The permutation also absorbs a bias specific to distinct trigrams: a trigram present on several chromosomes is counted
once in the whole proteome but once on each chromosome, and since shared trigrams are more often repeats,
chromosomes tend to have slightly fewer \ty{ABC} than the whole.

\subsection{Comparisons between genomes}\label{sec:between}

Whole-proteome pies of different organisms were compared by the same distance (Eq.~\ref{eq:js}), without a test,
since there is no common pool to permute. To rank genomes by how closely their chromosomes follow their own
whole-proteome pie, we took for each genome the median distance of its tested components, for distinct trigrams
(Table~\ref{tab:rank}). The two genomes with no components to test cannot be ranked.

\subsection{Organisms}

Table~\ref{tab:species} lists the 16 proteomes: two \org{E.~coli} strains, two yeasts, the ciliate
\org{Tetrahymena thermophila}, the nematode \org{C.~elegans}, the fruit fly, three plants and six vertebrates. The
units of comparison differ: the fly proteome is split by chromosome arm (2L, 2R, 3L, 3R, X, 4), the allotetraploid
\org{Xenopus laevis} by its L and S homeologous chromosomes \cite{session2016xenopus}, and \org{E.~coli} O157:H7 by replicon;
\org{E.~coli} K-12 (one replicon) and \org{Tetrahymena} (unassembled contigs only) have no components to compare. The current
reference proteomes contain no fly Y chromosome proteins and no \org{Arabidopsis} chloroplast proteins.

\begin{table}[t]
\centering
\caption{Proteomes and their trigram content. Entries: UniProtKB 2026\_04 entries in the reference proteome.
Excl.: entries newer than Pfam~39.0, not scanned by Pfam. Pfam prot.: proteins with at least one Pfam domain after
MD5 de-duplication. $\geq$3: proteins with at least three domains. Units: components tested against the whole
proteome (\S\ref{sec:sim}). Source: API = InterPro API; file = Pfam release file.}
\label{tab:species}
\footnotesize
\setlength{\tabcolsep}{2.9pt}
\begin{tabular}{l l r r r r r r r r l}
\toprule
Organism & Proteome & Entries & Excl. & Pfam prot. & $\geq$3 & Pfam fam. & Distinct & Occurr. & Units & Source\\
\midrule
\org{E.~coli} K-12        & UP000000625 &   4{,}403 &        -- &   4{,}164 &    538 & 3{,}390 &    724 &       943 &  -- & API\\
\org{E.~coli} O157:H7     & UP000000558 &   5{,}056 &        -- &   4{,}712 &    602 & 3{,}571 &    789 &    1{,}115 &   2 & API\\
\org{S.~pombe}            & UP000002485 &   5{,}130 &        -- &   4{,}577 &    796 & 3{,}946 & 1{,}093 &    1{,}741 &   3 & API\\
\org{S.~cerevisiae}       & UP000002311 &   6{,}067 &        -- &   5{,}377 &    895 & 4{,}383 & 1{,}071 &    1{,}748 &  16 & API\\
\org{T.~thermophila}      & UP000009168 &  26{,}878 &        -- &  11{,}971 & 2{,}520 & 3{,}259 & 1{,}252 &    7{,}565 &  -- & API\\
\org{C.~elegans}          & UP000001940 &  26{,}620 &        -- &  19{,}004 & 3{,}253 & 5{,}547 & 2{,}435 &   11{,}552 &   6 & API\\
\org{D.~melanogaster}     & UP000000803 &  21{,}941 &        -- &  17{,}481 & 4{,}540 & 6{,}325 & 3{,}107 &   18{,}023 &   6 & API\\
\org{A.~thaliana}         & UP000006548 &  39{,}144 &         1 &  33{,}237 & 6{,}105 & 5{,}850 & 2{,}388 &   14{,}632 &   5 & file\\
\org{O.~sativa}           & UP000059680 &  48{,}896 &         0 &  28{,}909 & 4{,}006 & 5{,}619 & 1{,}850 &    9{,}016 &  12 & file\\
\org{Z.~mays}             & UP000007305 &  63{,}237 &         0 &  49{,}478 & 7{,}373 & 5{,}778 & 2{,}468 &   15{,}454 &  11 & file\\
\org{D.~rerio}            & UP000000437 &  69{,}259 &        32 &  64{,}437 & 28{,}142 & 9{,}056 & 6{,}451 &  136{,}551 &  26 & file\\
\org{X.~laevis}           & UP000186698 &  62{,}048 &        22 &  57{,}898 & 22{,}020 & 9{,}218 & 6{,}595 &   89{,}336 &  19 & file\\
\org{M.~mulatta}          & UP000006718 &  44{,}389 &         0 &  40{,}943 & 14{,}865 & 9{,}737 & 7{,}039 &   52{,}335 &  23 & file\\
\org{R.~norvegicus}       & UP000002494 &  52{,}831 &        13 &  51{,}038 & 16{,}526 & 9{,}752 & 7{,}035 &   57{,}931 &  23 & file\\
\org{M.~musculus}         & UP000000589 & 228{,}903 & 174{,}060 &  45{,}071 & 10{,}767 & 9{,}769 & 6{,}461 &   37{,}965 &  22 & file\\
\org{H.~sapiens}          & UP000005640 & 263{,}886 & 116{,}508 & 125{,}247 & 38{,}846 & 10{,}043 & 9{,}087 & 116{,}751 &  25 & file\\
\bottomrule
\end{tabular}
\end{table}

\section{The fruit fly: one pie for every chromosome arm}\label{sec:fly}

The fly reference proteome (UP000000803) has 21{,}941 entries, of which 18{,}373 carry at least one of 6{,}330 Pfam
families (45{,}885 matches). After de-duplication, 4{,}540 proteins have three or more domains and yield 3{,}107
distinct trigrams with 18{,}023 occurrences.

\begin{figure}[p]
\centering
\includegraphics[width=\linewidth]{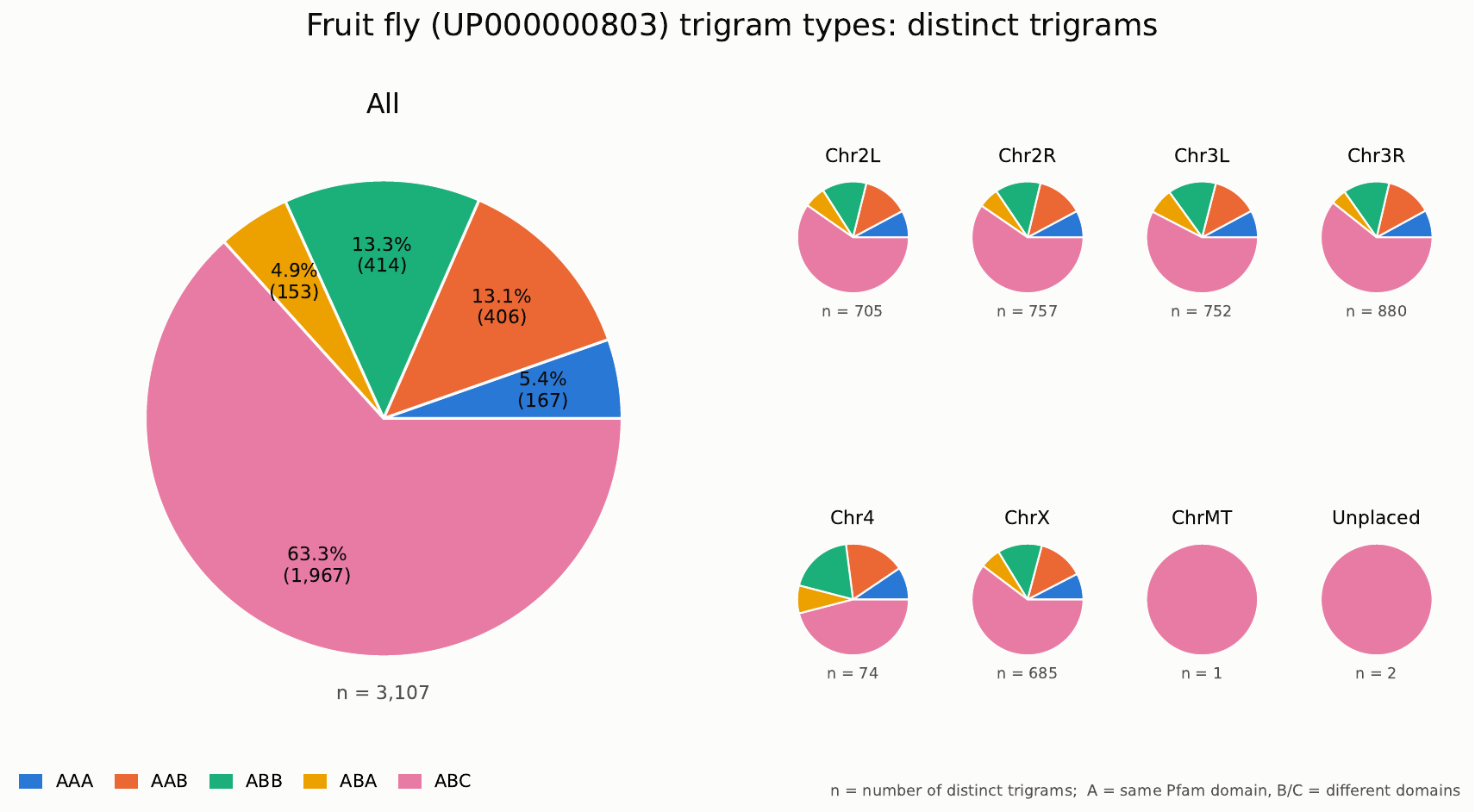}
\caption{Trigram types of the fly proteome, distinct trigrams. Large pie: whole proteome (percentages and counts);
small pies: each chromosome arm, the mitochondrial genome (ChrMT) and unplaced scaffolds. $n$: number of distinct
trigrams. Notebook \texttt{fruitfly\_02\_trigram\_analysis}.}
\label{fig:flypie}
\vspace{6mm}
\includegraphics[width=\linewidth]{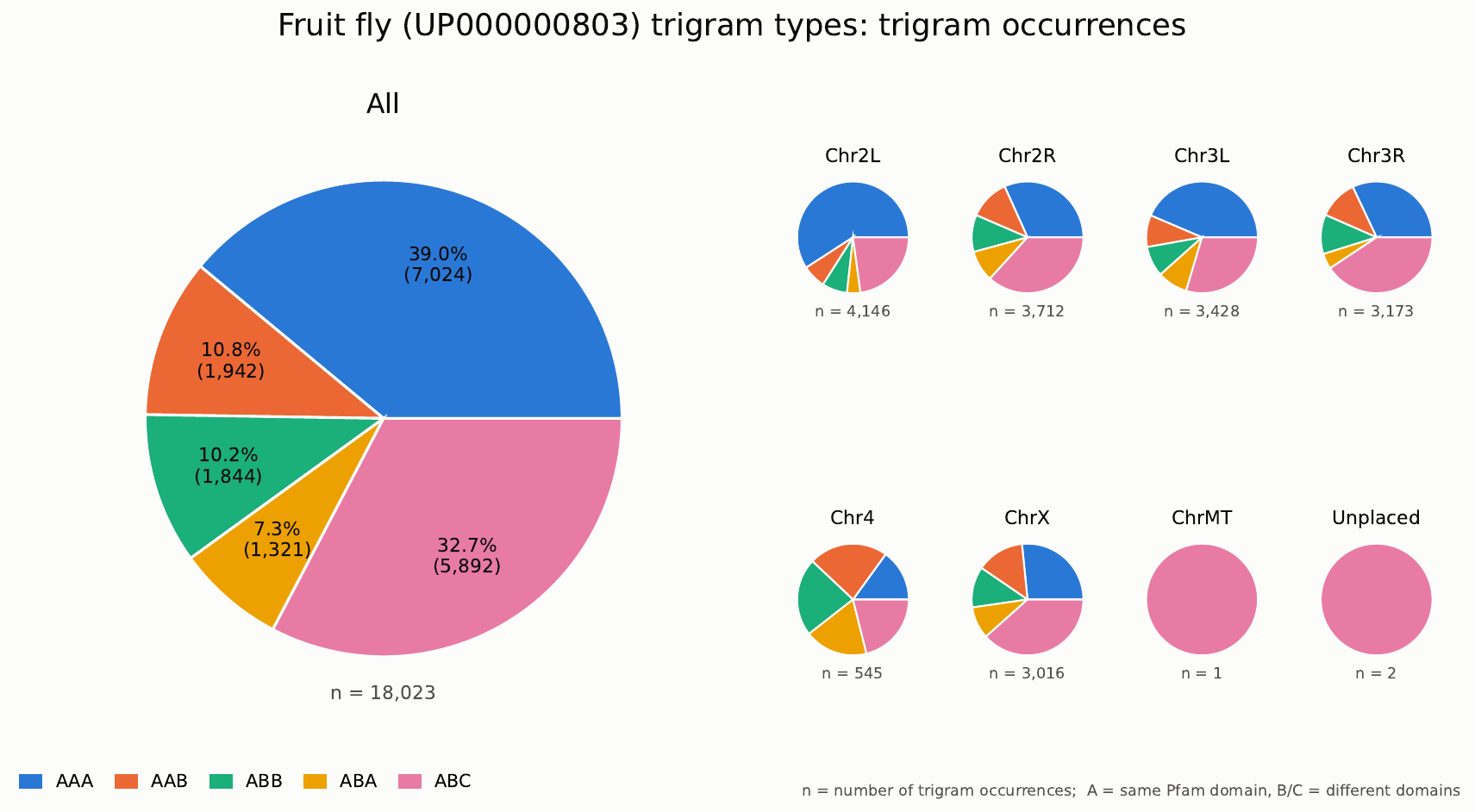}
\caption{As Fig.~\ref{fig:flypie}, for trigram occurrences.}
\label{fig:flypieTotal}
\end{figure}

\begin{figure}[t]
\centering
\includegraphics[width=\linewidth]{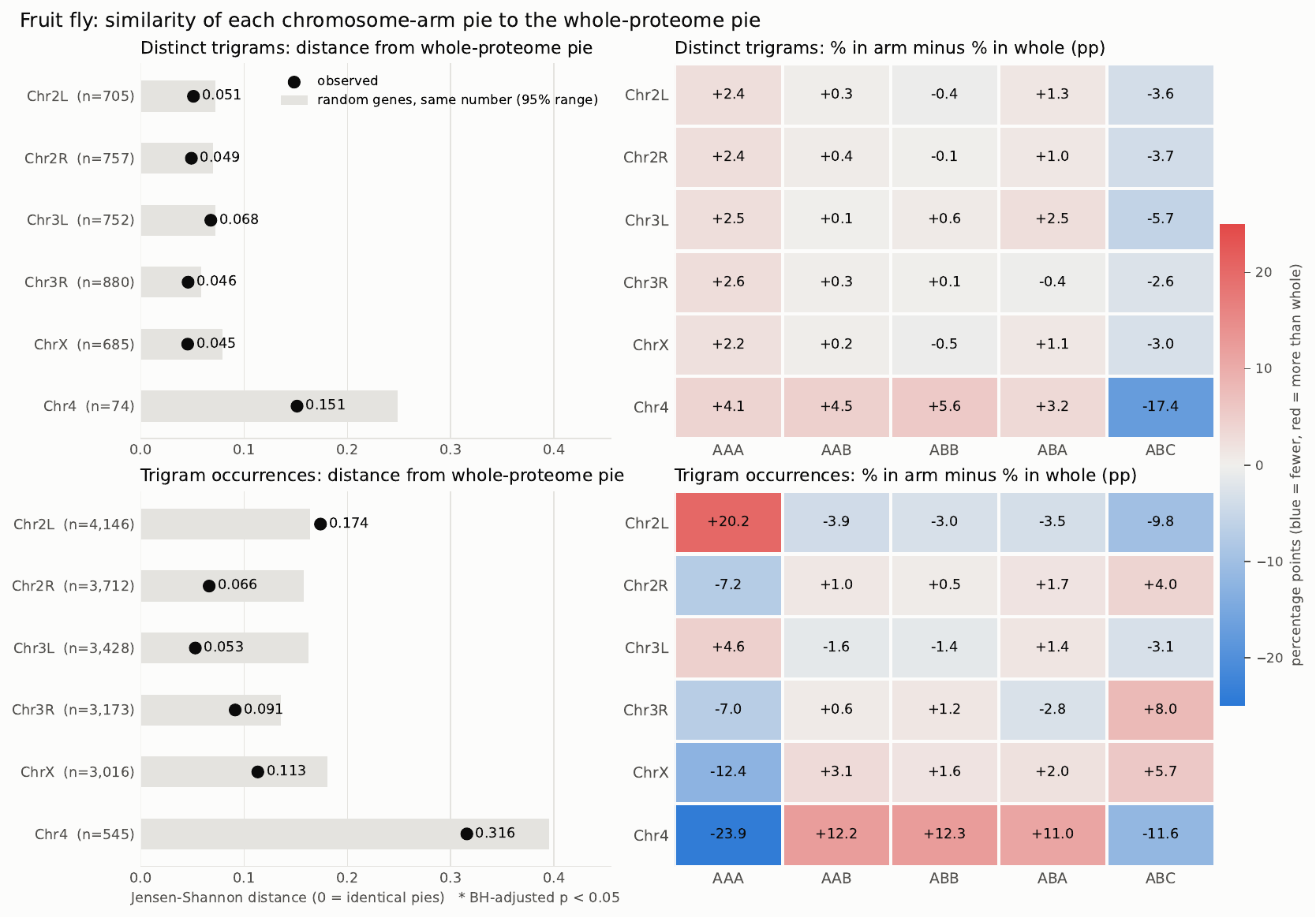}
\caption{Similarity of each fly arm to the whole proteome. Left: Jensen--Shannon distance of the arm pie from the
whole-proteome pie (dots) and the 95\% range of random gene sets of the same size (grey bars; 5{,}000 gene-level
permutations). Right: percentage of each type on the arm minus that in the whole proteome. Top: distinct trigrams;
bottom: trigram occurrences. No arm differs significantly (all BH $q\ge0.19$).}
\label{fig:flysim}
\end{figure}

\paragraph{Distinct trigrams.} Almost two thirds of the distinct trigrams are \ty{ABC} (63.3\%), followed by \ty{ABB}
(13.3\%) and \ty{AAB} (13.1\%), with \ty{AAA} (5.4\%) and \ty{ABA} (4.9\%) the rarest (Fig.~\ref{fig:flypie}). The six
arms reproduce this pie closely (Fig.~\ref{fig:flysim}): the five large arms are 0.045--0.068 from the whole proteome
(similarity 0.93--0.95), and the small chromosome~4, with only 74 distinct trigrams, is 0.151. All six distances lie
inside the range of random gene sets of the same size (BH $q\ge0.56$). The only consistent pattern, a deficit of \ty{ABC} on every arm
(3--6~pp on the large arms and 17~pp on chromosome~4), is the union effect of \S\ref{sec:sim}, not a property of any arm.

\paragraph{Trigram occurrences.} Counting occurrences changes the picture (Fig.~\ref{fig:flypieTotal}). The whole
proteome is now 39.0\% \ty{AAA} and 32.7\% \ty{ABC}, and the arms look more different from each other: arm 2L has 20~pp
more \ty{AAA} than the whole, chromosome~4 has 24~pp less, and X 12~pp less. Yet after correction no arm differs from
the whole more than a random gene set does (lowest: 2L, $p=0.031$, BH $q=0.19$). The reason is that a handful of
repeat-rich genes, each encoded by many isoforms, dominate the occurrences of their arm (Table~\ref{tab:flygenes}):
\textit{dumpy}, the gigantic extracellular matrix protein of the epidermis \cite{wilkin2000dumpy}, contributes 34\% of
the occurrences on 2L (16 isoforms, 1{,}408 occurrences, 1{,}350 of them \ty{AAA}), \textit{bt} 66\% of chromosome~4,
\textit{trol} 29\% of X, \textit{Dscam1} with its many splice isoforms \cite{schmucker2000dscam} 23\% of 2R, and
\textit{sls} 20\% of 3L. An occurrence pie is therefore a statement about a few genes; the distinct-trigram pie
describes the arm.

\begin{figure}[t]
\centering
\includegraphics[width=0.85\linewidth]{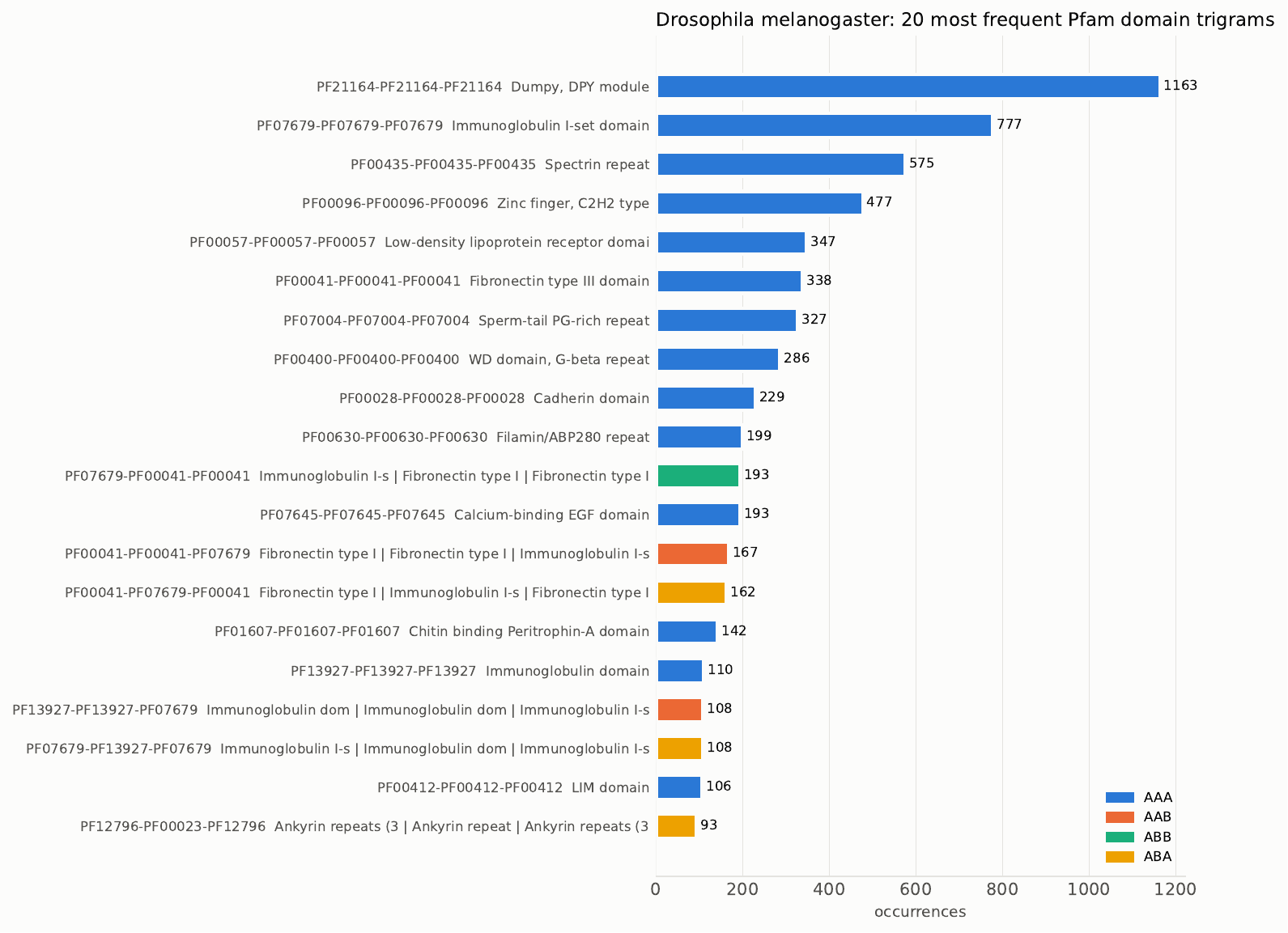}
\caption{The 20 most frequent trigrams of the fly proteome, coloured by type.}
\label{fig:flytop}
\end{figure}

\paragraph{Most frequent trigrams.} The commonest trigrams of each type (Table~\ref{tab:toptrigrams}; Fig.~\ref{fig:flytop}) are built from a
small set of extracellular and structural repeat domains: the Dumpy DPY module (1{,}163 occurrences), immunoglobulin
I-set (777), spectrin (575), C2H2 zinc finger (477) and LDL receptor class~A (347) repeats for \ty{AAA}; and combinations of
immunoglobulin and fibronectin type~III domains for \ty{AAB}, \ty{ABB}, \ty{ABA} and \ty{ABC}.

\begin{table}[t]
\centering
\caption{The gene carrying the most trigram occurrences on each fly arm. Iso.: isoforms (proteins with $\geq$3 domains)
in the de-duplicated proteome. \% arm: share of the arm's trigram occurrences.}
\label{tab:flygenes}
\small
\begin{tabular}{l l r r r r}
\toprule
Arm & Gene & Iso. & Occurrences & of which \ty{AAA} & \% arm\\
\midrule
2L & \textit{dpy}    & 16 & 1{,}408 & 1{,}350 & 34.0\\
2R & \textit{Dscam1} & 62 &     858 &     164 & 23.1\\
3L & \textit{sls}    & 13 &     684 &     588 & 20.0\\
3R & \textit{cher}   &  9 &     149 &     137 &  4.7\\
X  & \textit{trol}   & 25 &     865 &     250 & 28.7\\
4  & \textit{bt}     &  5 &     359 &      64 & 65.9\\
\bottomrule
\end{tabular}
\end{table}

\begin{table}[t]
\centering
\caption{The three most frequent trigrams of each type in the whole fly and human proteomes (occurrences). Names are
abbreviated from Pfam.}
\label{tab:toptrigrams}
\footnotesize
\setlength{\tabcolsep}{3.5pt}
\begin{tabular}{l l r l r}
\toprule
& \multicolumn{2}{c}{\org{D.~melanogaster}} & \multicolumn{2}{c}{\org{H.~sapiens}}\\
\cmidrule(lr){2-3}\cmidrule(lr){4-5}
Type & Trigram & $n$ & Trigram & $n$\\
\midrule
\multirow{3}{*}{\ty{AAA}} & DPY module $\times3$ & 1{,}163 & zf-C2H2 $\times3$ & 8{,}753\\
 & Ig I-set $\times3$ & 777 & WD40 $\times3$ & 2{,}633\\
 & Spectrin $\times3$ & 575 & Ankyrin (3 copies) $\times3$ & 1{,}860\\
\addlinespace
\multirow{3}{*}{\ty{AAB}} & FN3--FN3--Ig I-set & 167 & FN3--FN3--Ig I-set & 376\\
 & Ig--Ig--Ig I-set & 108 & zf-C2H2--zf-C2H2--zf-H2C2 & 183\\
 & Ig I-set--Ig I-set--FN3 & 51 & Ig I-set--Ig I-set--FN3 & 142\\
\addlinespace
\multirow{3}{*}{\ty{ABB}} & Ig I-set--FN3--FN3 & 193 & KRAB--zf-C2H2--zf-C2H2 & 661\\
 & Ig I-set--Ig--Ig & 86 & Ig I-set--FN3--FN3 & 491\\
 & zf-AD--zf-C2H2--zf-C2H2 & 63 & Pumilio N-term.--PUF--PUF & 208\\
\addlinespace
\multirow{3}{*}{\ty{ABA}} & FN3--Ig I-set--FN3 & 162 & FN3--Ig I-set--FN3 & 365\\
 & Ig I-set--Ig--Ig I-set & 108 & cbEGF--TB--cbEGF & 166\\
 & Ank (3 copies)--Ank--Ank (3 copies) & 93 & zf-C2H2--zf-H2C2--zf-C2H2 & 164\\
\addlinespace
\multirow{3}{*}{\ty{ABC}} & Ig--Ig I-set--FN3 & 89 & FERM N--FERM M--FERM C & 276\\
 & Ig I-set--FN3--DSCAM Fn3-like & 77 & FERM M--FERM C--FERM adjacent & 239\\
 & FN3--DSCAM Fn3-like--DSCAM C & 62 & Acyl-CoA DH N--M--C & 161\\
\bottomrule
\end{tabular}
\end{table}

\FloatBarrier
\section{Human: one chromosome stands out}\label{sec:human}

The human reference proteome (UP000005640) is the largest in this study. Of its 263{,}886 entries, 116{,}508 have
sequences newer than Pfam~39.0 and were excluded, leaving 147{,}378 entries from 21{,}511 genes. After de-duplication,
125{,}247 proteins carry at least one of 10{,}043 Pfam families, and 38{,}846 proteins have three or more domains. These
yield 9{,}087 distinct trigrams with 116{,}751 occurrences.

\begin{figure}[p]
\centering
\includegraphics[width=0.82\linewidth]{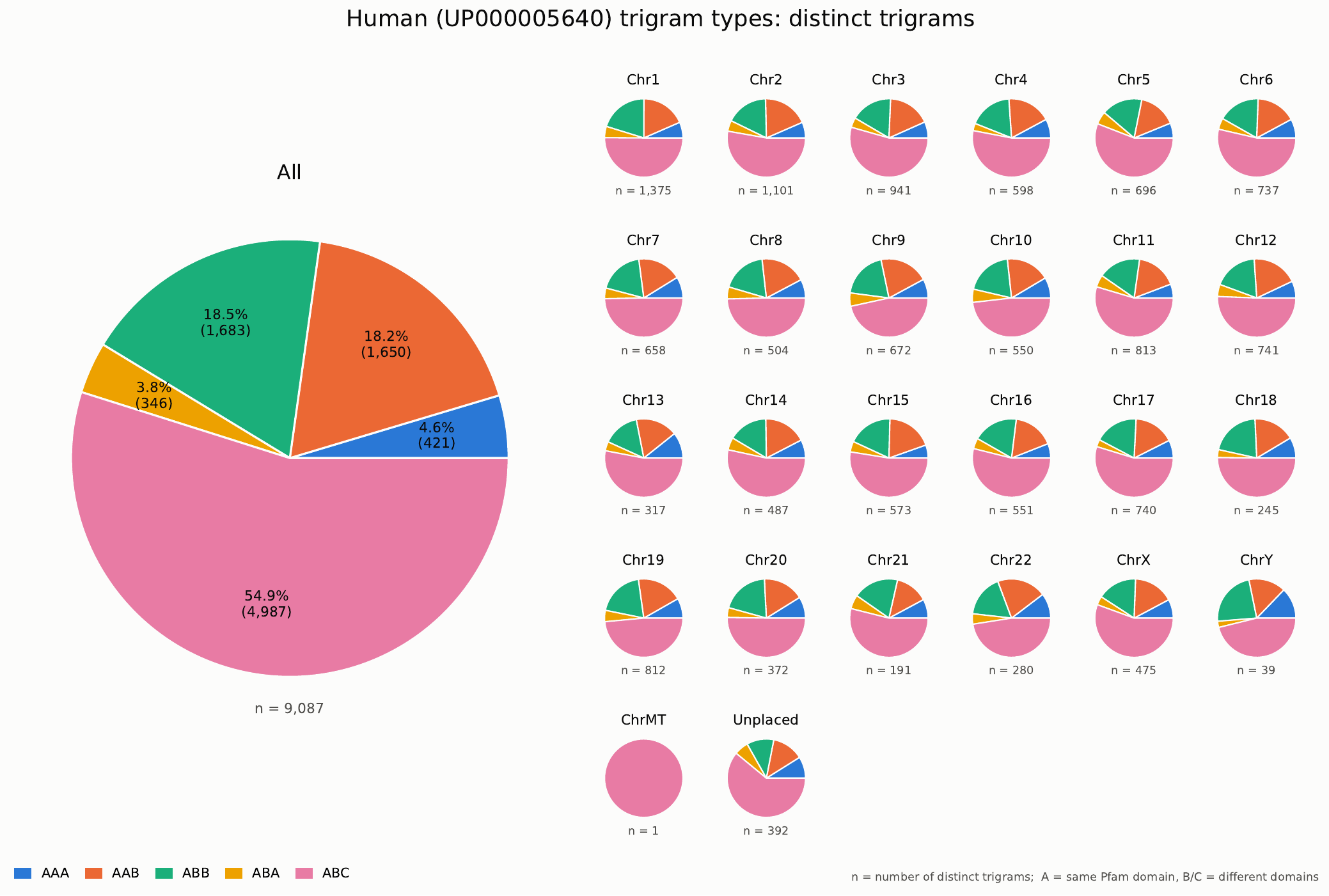}
\caption{Human trigram types, distinct trigrams: whole proteome (large pie) and each chromosome. Notebook
\texttt{human\_trigram}.}
\label{fig:humanpie}
\vspace{5mm}
\includegraphics[width=0.82\linewidth]{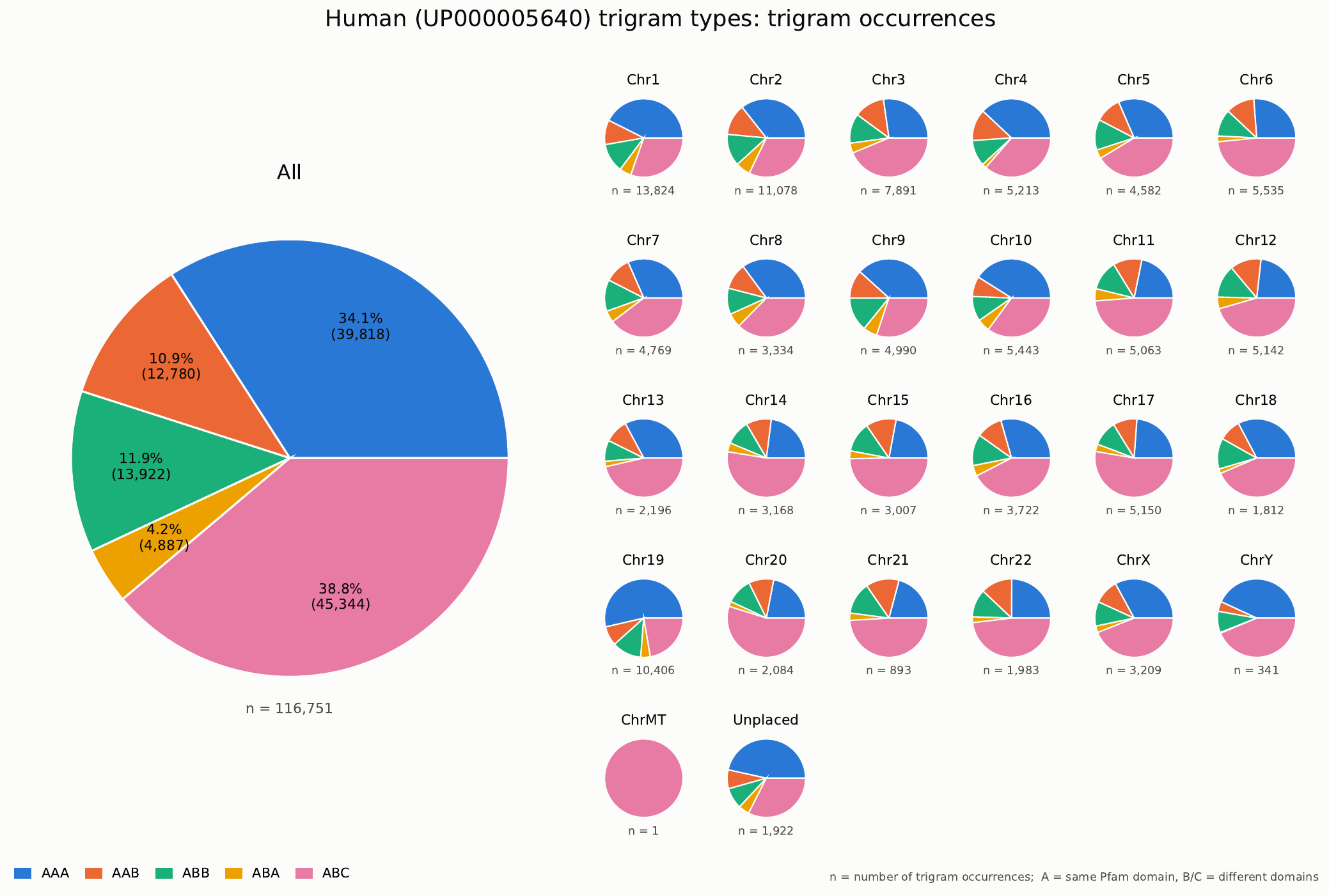}
\caption{As Fig.~\ref{fig:humanpie}, for trigram occurrences. Chromosome~19 is the most \ty{AAA}-rich chromosome.}
\label{fig:humanpieTotal}
\end{figure}

\begin{figure}[p]
\centering
\includegraphics[height=0.88\textheight]{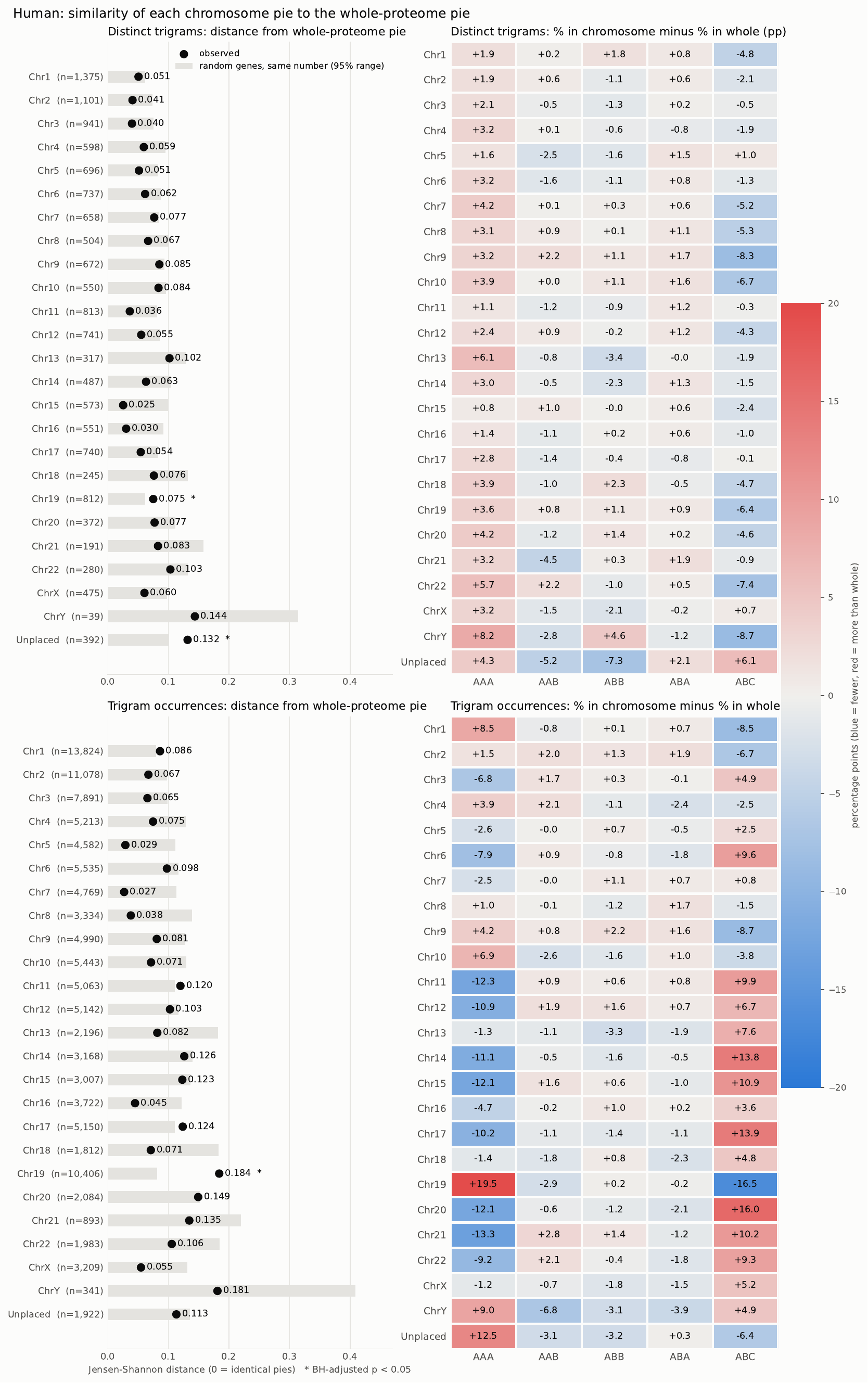}
\caption{Human: similarity of each chromosome to the whole proteome, drawn as Fig.~\ref{fig:flysim}.}
\label{fig:humansim}
\end{figure}

\paragraph{Distinct trigrams.} The human pie has the lowest \ty{ABC} share of the 16 organisms (54.9\%). \ty{ABB} (18.5\%) and
\ty{AAB} (18.2\%) are each about 5~pp higher than in the fly, while \ty{AAA} (4.6\%) and \ty{ABA} (3.8\%) remain the rarest
(Fig.~\ref{fig:humanpie}); the distance between the human and fly pies is 0.095. Of the 25 tested components (22 autosomes,
X, Y and the unplaced proteins; Fig.~\ref{fig:humansim}), 23 lie inside the random-gene range. The median distance from the
whole proteome is 0.063; the autosomes range from 0.025 (chromosome~15) to 0.103 (chromosome~22), and the Y chromosome, with
13 genes and 39 distinct trigrams, is at 0.144. Two components differ after correction: chromosome~19 (distance 0.075,
\ty{ABC} $-$6.4~pp, $q=0.025$) and the unplaced proteins (0.132, \ty{ABB} $-$7.3~pp, $q=0.020$).

\paragraph{Trigram occurrences.} The occurrence pie is 38.8\% \ty{ABC} and 34.1\% \ty{AAA}, with 11.9\% \ty{ABB}, 11.0\%
\ty{AAB} and 4.2\% \ty{ABA} (Fig.~\ref{fig:humanpieTotal}). Distances run from 0.027 (chromosome~7) to 0.184 (chromosome~19),
and 21 of the 25 components are inside the random-gene range. Only chromosome~19 differs after correction (\ty{AAA}
+19.5~pp, $q=0.005$): 53.6\% of its occurrences are \ty{AAA}, the highest share of any human chromosome.

Chromosome~19 has the highest gene density of all human chromosomes and many tandemly arranged gene families
\cite{grimwood2004chr19}, including the largest clusters of KRAB zinc-finger genes \cite{huntley2006krab}. Triples of the C2H2
zinc finger (\pf{PF00096}) alone give 4{,}370 of its 10{,}406 occurrences (42\%) and half of the 8{,}753 such triples in the
whole proteome. The excess is spread over many genes: the largest single contributor on chromosome~19, \textit{LTBP4}, gives
only 3.6\% of its occurrences. Elsewhere one gene can dominate a chromosome, as in the fly: \textit{TTN}, which encodes titin
\cite{labeit1995titin}, gives 19\% of the occurrences on chromosome~2, \textit{NRAP} 21\% on chromosome~10 (74 isoforms),
\textit{ANK2} 18\% on chromosome~4 (79 isoforms) and \textit{PUM1} 10\% on chromosome~1 (188 isoforms). None of these
chromosomes differs at the gene level.

\begin{figure}[t]
\centering
\includegraphics[width=0.85\linewidth]{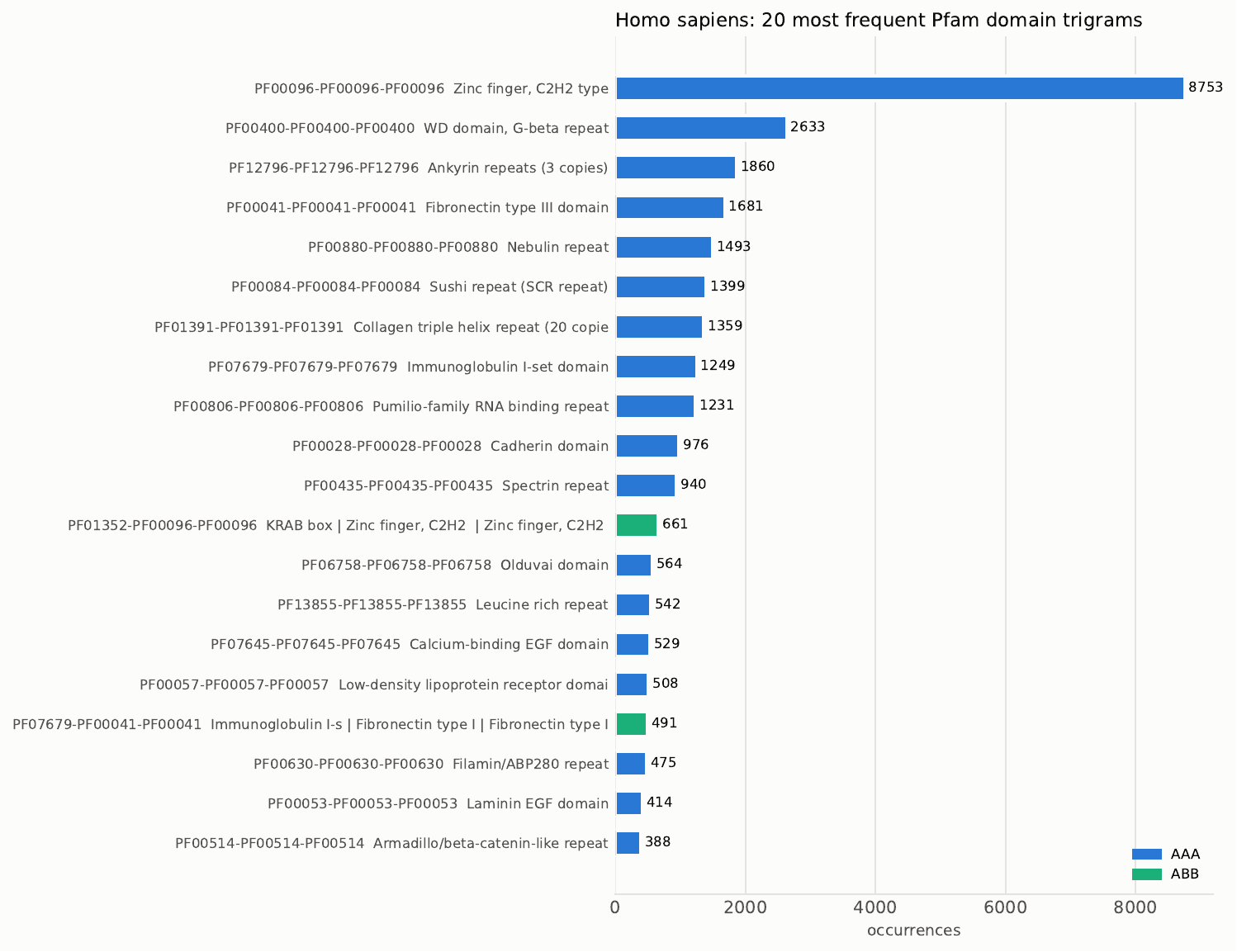}
\caption{The 20 most frequent trigrams of the human proteome, coloured by type.}
\label{fig:humantop}
\end{figure}

\paragraph{Most frequent trigrams.} The commonest human trigrams (Table~\ref{tab:toptrigrams}; Fig.~\ref{fig:humantop}) are
triples of C2H2 zinc finger (8{,}753), WD40 (2{,}633) and ankyrin repeats (1{,}860), and KRAB--zf-C2H2--zf-C2H2 is the commonest
\ty{ABB}. The immunoglobulin and fibronectin combinations that lead the fly list recur among the leading human \ty{AAB},
\ty{ABB} and \ty{ABA} trigrams.

\FloatBarrier
\section{Overall results for the 16 organisms}\label{sec:summary}

\subsection{Whole-proteome pies}\label{sec:species}

\begin{figure}[t]
\centering
\includegraphics[width=\linewidth]{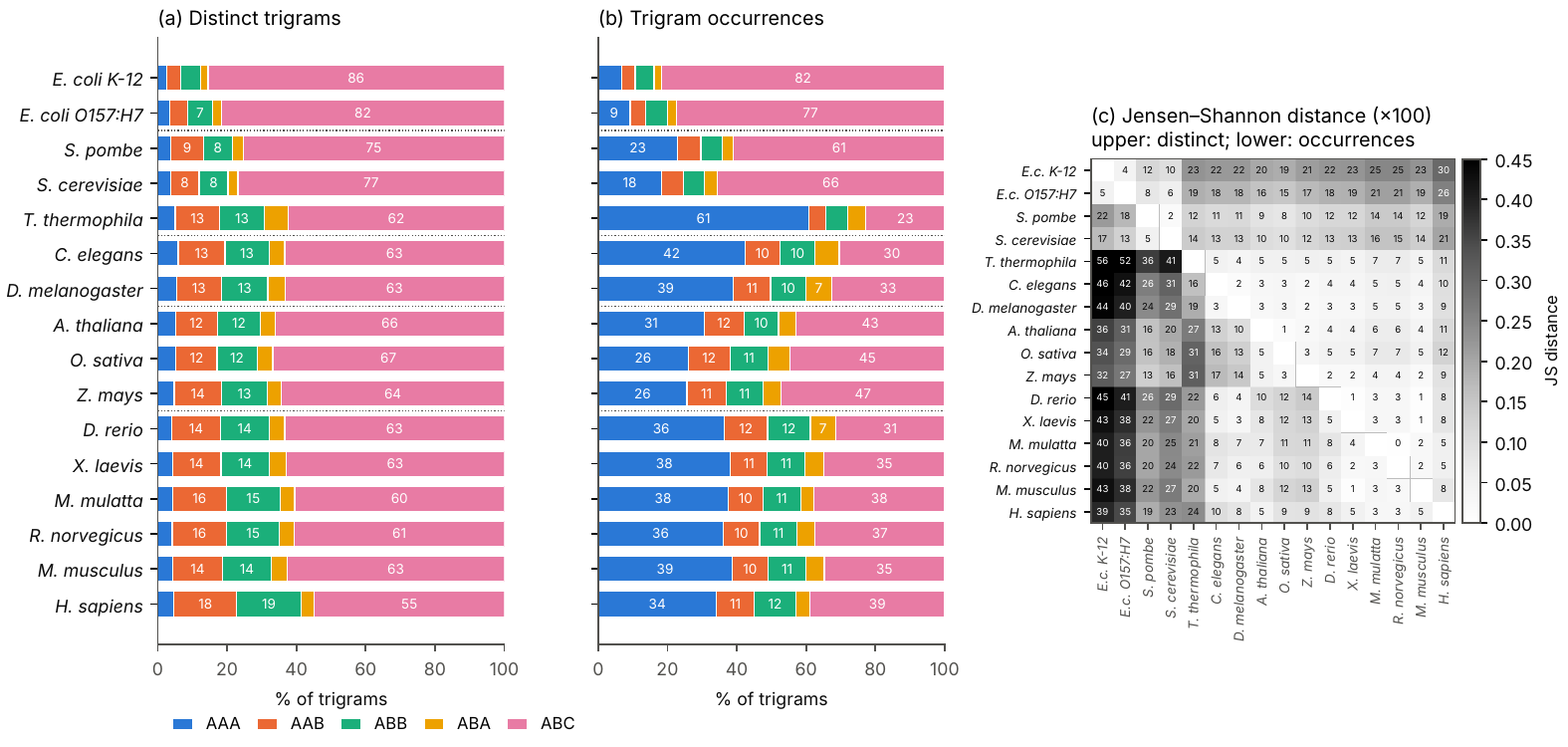}
\caption{Whole-proteome trigram types of 16 organisms, ordered bacteria, yeasts and ciliate, invertebrates, plants,
vertebrates (dotted lines). (a) Distinct trigrams and (b) trigram occurrences, as percentages (numbers $\geq$7\% shown);
(c) Jensen--Shannon distance ($\times100$) between every pair of whole-proteome pies, distinct trigrams above the
diagonal and occurrences below. Computed from the \texttt{<species>\_trigram\_summary.csv} tables
(\texttt{paper/scripts/make\_crossspecies\_figures.py}).}
\label{fig:species}
\end{figure}

Figure~\ref{fig:species} and Table~\ref{tab:pct} set the whole-proteome pies of the 16 organisms side by side.

\paragraph{Distinct trigrams.} The pies of the eleven multicellular eukaryotes are close to one another: \ty{ABC}
55--67\%, \ty{AAB} 12--18\%, \ty{ABB} 12--19\%, \ty{AAA} 4--6\% and \ty{ABA} 4--5\%, with a median pairwise distance of
0.036 (range 0.003--0.124) and 0.028 among the six vertebrates. Human is the most distant of the eleven (0.054--0.124 from the
others). These distances are about half the median distance (0.070) between a single chromosome and its own proteome,
although that comparison favours whole proteomes, which contain many more trigrams. \org{Tetrahymena} (62\% \ty{ABC}) is
0.036--0.066 from the multicellular organisms other than human (0.110). The yeasts (75--77\% \ty{ABC}, 8--9\% \ty{AAB}/\ty{ABB})
and the bacteria (82--86\% \ty{ABC}, 4--7\% \ty{AAB}/\ty{ABB}) are progressively further away (up to 0.30 between
\org{E.~coli} K-12 and human), and are close to each other within each pair (0.019 between the yeasts, 0.045 between the
\org{E.~coli} strains). Multicellular eukaryotes are at least 0.080 from either yeast and at least 0.148 from either bacterium.

\paragraph{\ty{AAB} equals \ty{ABB}.} In every organism the shares of \ty{AAB} and \ty{ABB} are within 2~pp of each other,
for distinct trigrams ($-1.9$ to $+1.0$~pp) and for occurrences ($-1.8$ to $+1.6$~pp). For occurrences this is largely a
combinatorial necessity: a run of identical domains that is flanked by other domains on both sides begins with one
\ty{ABB} and ends with one \ty{AAB} (Fig.~\ref{fig:workflow}b), so the two can differ only through runs that start or end
the protein. The near-equality for distinct trigrams is less obvious. It means that the vocabulary of domains that precede a
repeat run is as large as the vocabulary of those that follow it, and it holds from bacteria to human.

\paragraph{Trigram occurrences.} Occurrence pies vary more (median pairwise distance 0.074 among multicellular
eukaryotes, 0.16 overall), mainly in their share of \ty{AAA}: 7--9\% in bacteria, 18--23\% in yeasts, 26--31\% in plants,
34--42\% in animals, and 61\% in \org{Tetrahymena}, whose leucine-rich repeat proteins (\pf{PF13516}) alone give
2{,}957 occurrences. Like the per-chromosome occurrence pies, these reflect how many very large repeat proteins and
isoforms a proteome contains (titin \cite{labeit1995titin}, nebulin \cite{labeit1995nebulin} and C2H2 zinc-finger proteins in
vertebrates; Table~\ref{tab:toptrigrams}), and they are the part of the analysis most sensitive to annotation
(\S\ref{sec:caveats}).

\begin{table}[t]
\centering
\caption{Whole-proteome proportions (\%) of the five trigram types.}
\label{tab:pct}
\small
\setlength{\tabcolsep}{4.5pt}
\begin{tabular}{l rrrrr c rrrrr}
\toprule
& \multicolumn{5}{c}{Distinct trigrams} && \multicolumn{5}{c}{Trigram occurrences}\\
\cmidrule(lr){2-6}\cmidrule(lr){8-12}
Organism & \ty{AAA} & \ty{AAB} & \ty{ABB} & \ty{ABA} & \ty{ABC} && \ty{AAA} & \ty{AAB} & \ty{ABB} & \ty{ABA} & \ty{ABC}\\
\midrule
\org{E.~coli} K-12    & 2.5 &  4.1 &  5.7 & 2.2 & 85.5 &&  6.9 &  3.8 &  5.4 & 2.1 & 81.8\\
\org{E.~coli} O157:H7 & 3.4 &  5.2 &  7.1 & 2.7 & 81.6 &&  9.2 &  4.5 &  6.3 & 2.5 & 77.5\\
\org{S.~pombe}        & 3.8 &  9.3 &  8.3 & 3.2 & 75.3 && 22.9 &  6.7 &  6.2 & 3.3 & 60.9\\
\org{S.~cerevisiae}   & 3.6 &  8.3 &  8.3 & 2.9 & 76.8 && 18.3 &  6.4 &  6.1 & 3.6 & 65.7\\
\org{T.~thermophila}  & 5.0 & 12.8 & 12.9 & 6.9 & 62.4 && 60.8 &  4.9 &  6.4 & 5.2 & 22.7\\
\org{C.~elegans}      & 5.9 & 13.5 & 12.7 & 4.5 & 63.4 && 42.3 & 10.2 & 10.0 & 7.0 & 30.5\\
\org{D.~melanogaster} & 5.4 & 13.1 & 13.3 & 4.9 & 63.3 && 39.0 & 10.8 & 10.2 & 7.3 & 32.7\\
\org{A.~thaliana}     & 5.1 & 12.1 & 12.5 & 4.2 & 66.1 && 30.6 & 11.5 & 10.0 & 5.0 & 43.0\\
\org{O.~sativa}       & 5.2 & 11.9 & 11.5 & 4.5 & 66.8 && 26.0 & 12.2 & 10.8 & 6.4 & 44.6\\
\org{Z.~mays}         & 4.8 & 13.6 & 13.3 & 4.1 & 64.3 && 25.6 & 11.4 & 10.6 & 5.3 & 47.2\\
\org{D.~rerio}        & 4.1 & 14.0 & 14.1 & 4.5 & 63.4 && 36.4 & 12.5 & 12.4 & 7.3 & 31.5\\
\org{X.~laevis}       & 4.2 & 14.0 & 14.0 & 4.8 & 63.0 && 38.2 & 10.6 & 10.8 & 5.6 & 34.8\\
\org{M.~mulatta}      & 4.3 & 15.6 & 15.4 & 4.3 & 60.5 && 37.5 & 10.0 & 10.9 & 3.9 & 37.7\\
\org{R.~norvegicus}   & 4.2 & 15.7 & 15.3 & 4.3 & 60.5 && 36.1 & 10.5 & 10.9 & 5.1 & 37.4\\
\org{M.~musculus}     & 4.3 & 14.4 & 14.1 & 4.6 & 62.6 && 38.7 & 10.4 & 11.0 & 5.2 & 34.8\\
\org{H.~sapiens}      & 4.6 & 18.2 & 18.5 & 3.8 & 54.9 && 34.1 & 11.0 & 11.9 & 4.2 & 38.8\\
\bottomrule
\end{tabular}
\end{table}

\subsection{Chromosomes against their own proteome}\label{sec:chrom}

\begin{figure}[t]
\centering
\includegraphics[width=\linewidth]{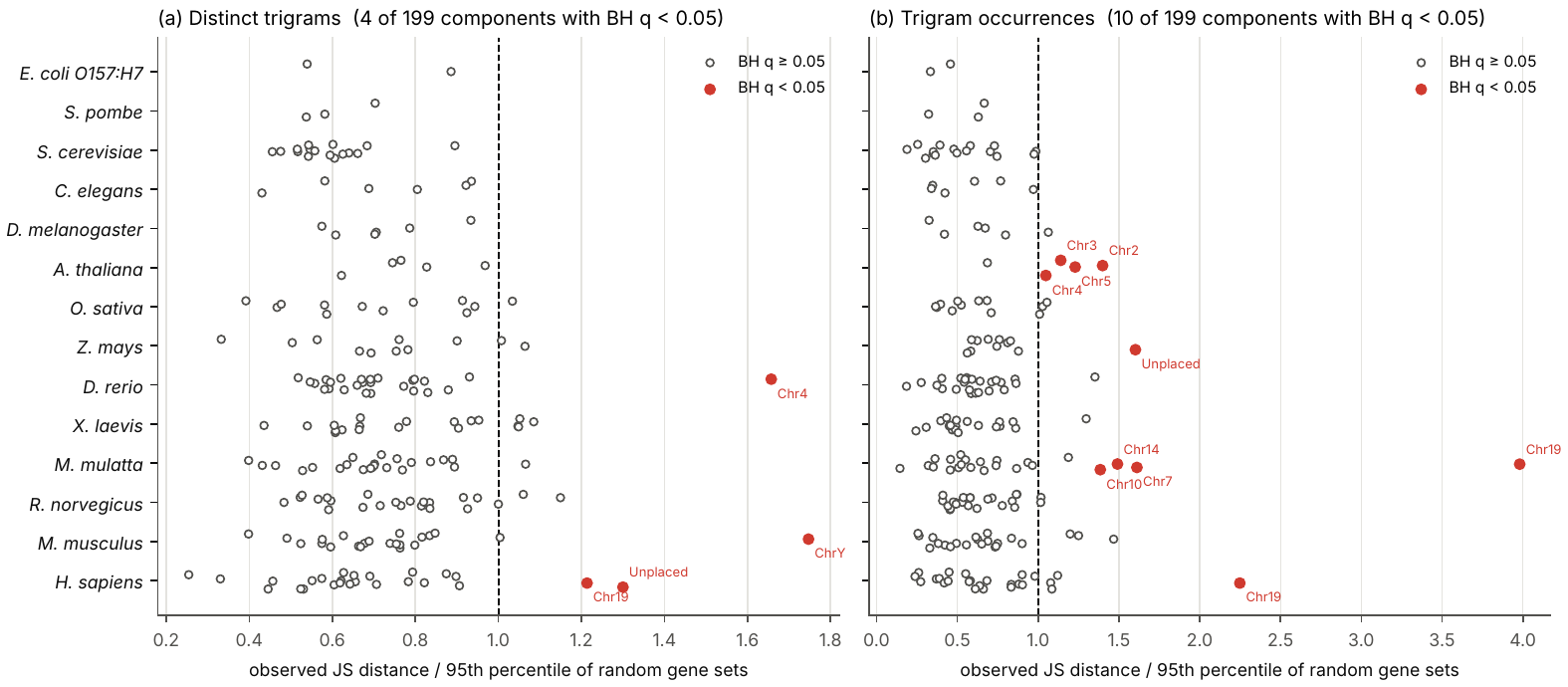}
\caption{Every tested component of 14 organisms against its whole proteome. Each dot is one chromosome (or arm,
replicon, organelle or unplaced set), placed at its Jensen--Shannon distance from the whole-proteome pie divided by
the 95th percentile of the distance for random gene sets of the same size; values left of the dashed line lie inside the
random-gene range. Red: BH-adjusted $q<0.05$ (Table~\ref{tab:sig}). (a) Distinct trigrams; (b) trigram occurrences.
Vertical jitter is for visibility only.}
\label{fig:compsim}
\end{figure}

The test of \S\ref{sec:sim} was run on every component with at least ten genes in the 14 organisms that have more than one
(Fig.~\ref{fig:compsim}): 199 components, from the 2 replicons of \org{E.~coli} O157:H7 to the 26 chromosomes of zebrafish.
For distinct trigrams, the median distance from the whole proteome is 0.070, and 184 of the 199 components (92\%) lie
inside the random-gene range; after correction, 4 (2.0\%) differ. For occurrences the median is 0.075, 174 components
(87\%) are inside the range, and 10 (5.0\%) differ. The 14 significant cases (Table~\ref{tab:sig}) fall into three groups.

\begin{figure}[p]
\centering
\includegraphics[height=0.88\textheight]{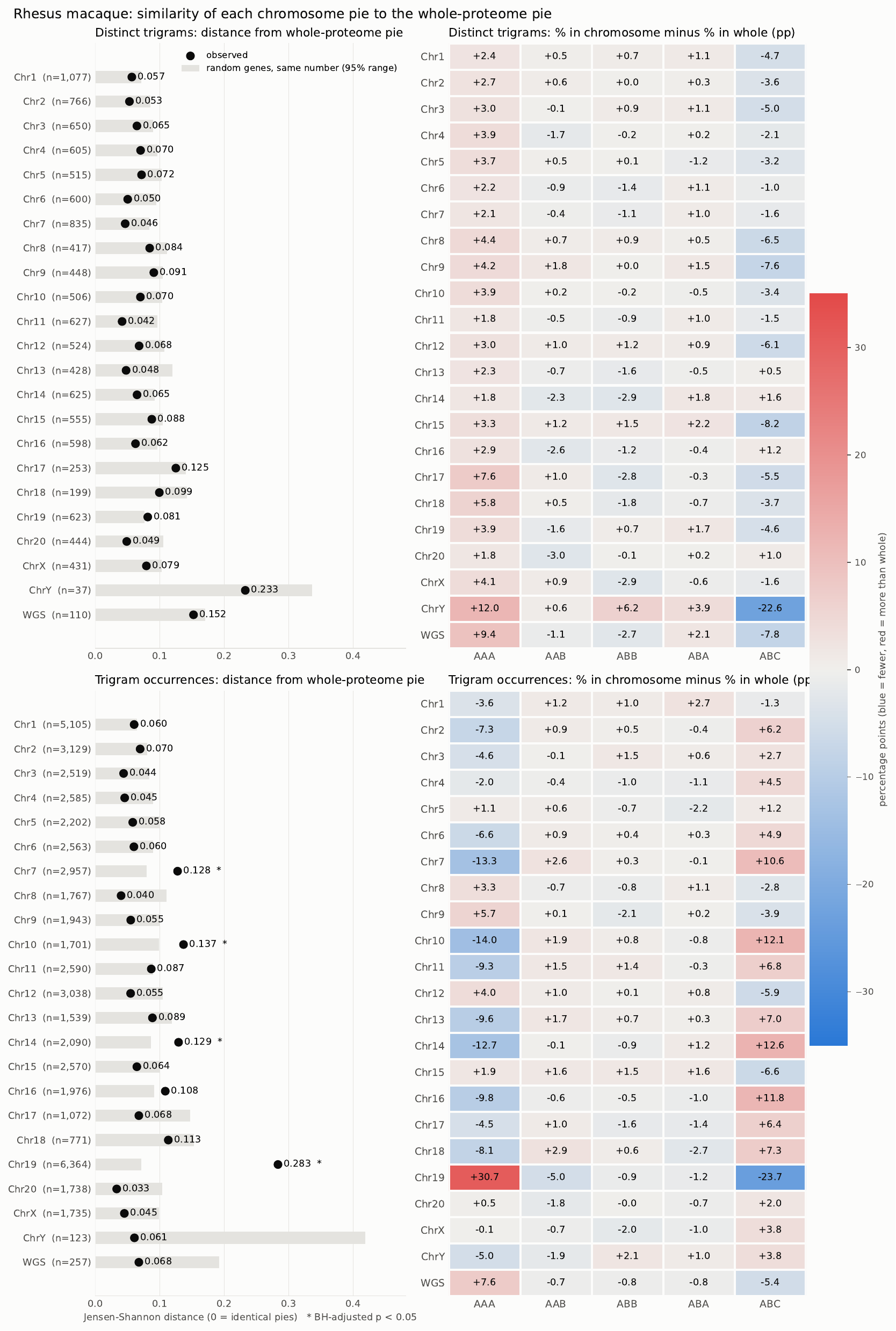}
\caption{Rhesus macaque: similarity of each chromosome to the whole proteome, drawn as Fig.~\ref{fig:flysim}.
Chromosome~19 has the largest \ty{AAA} excess of any component in this study.}
\label{fig:macsim}
\end{figure}

\paragraph{Tandem gene clusters.} The strongest deviation in the whole study is macaque chromosome~19, whose occurrence
pie has 30.7~pp more \ty{AAA} than the macaque proteome ($q=0.005$; Fig.~\ref{fig:macsim}). Triples of C2H2 zinc finger give
3{,}952 of its 6{,}364 occurrences (62\%), as on human chromosome~19 (\S\ref{sec:human}). The macaque chromosomes 7, 10 and 14
that are \ty{AAA}-poor (12.7--14.0~pp less) are probably in part the counterpart of this excess, which raises the \ty{AAA}
share of the whole proteome. Zebrafish chromosome~4, whose long arm carries clusters of multi-zinc-finger and NLR genes
\cite{howe2013zebrafish,howe2016nlr}, has 9.8~pp less \ty{ABC} among its distinct trigrams ($q=0.005$).

\paragraph{Unusual components.} Mouse chromosome~Y (46 genes, 20 distinct trigrams) has 22~pp more \ty{ABC}; the unplaced
maize proteins (58 genes), 29~pp more \ty{ABC} in occurrences; and the human unplaced proteins, 7~pp less \ty{ABB}. These
are small or heterogeneous sets rather than chromosomes in the usual sense.

\paragraph{Small but consistent shifts.} Four of the five \org{Arabidopsis} chromosomes differ in occurrences, with
distances of only 0.04--0.07 (\S\ref{sec:most}).

Everywhere else, including all 23 rat, 19 \org{Xenopus}, 16 budding-yeast, 12 rice and 6 \org{C.~elegans} components, no
component differs. Where occurrence pies of single chromosomes look very different, one gene usually accounts for it:
\textit{obscnb}, an obscurin gene \cite{young2001obscurin}, gives 48\% of the occurrences of zebrafish chromosome~24, the titin
gene \textit{ttn.1} 47\% of zebrafish chromosome~9, and \textit{Neb} \cite{labeit1995nebulin} 18\% of rat chromosome~3; none is
significant at the gene level.

\begin{table}[t]
\centering
\caption{Components whose pie differs from their whole proteome (BH $q<0.05$). Genes: genes with $\geq$3-domain
proteins; $n$: trigrams; JS: distance from the whole; $q_{95}$: 95th percentile for random gene sets; largest
difference in pp.}
\label{tab:sig}
\small
\begin{tabular}{l l l r r r r l r}
\toprule
Organism & Measure & Component & Genes & $n$ & JS & $q_{95}$ & Largest diff. & $q$\\
\midrule
\org{H.~sapiens}   & distinct    & Chr19    & 628 &    812 & 0.075 & 0.062 & \ty{ABC} $-$6.4  & 0.025\\
\org{H.~sapiens}   & distinct    & Unplaced & 226 &    392 & 0.132 & 0.101 & \ty{ABB} $-$7.3  & 0.020\\
\org{M.~musculus}  & distinct    & ChrY     &  46 &     20 & 0.363 & 0.208 & \ty{ABC} +22.4 & 0.004\\
\org{D.~rerio}     & distinct    & Chr4     & 1{,}038 & 472 & 0.105 & 0.064 & \ty{ABC} $-$9.8  & 0.005\\
\addlinespace
\org{H.~sapiens}   & occurrences & Chr19    & 628 & 10{,}406 & 0.184 & 0.082 & \ty{AAA} +19.5 & 0.005\\
\org{M.~mulatta}   & occurrences & Chr19    & 504 &  6{,}364 & 0.283 & 0.071 & \ty{AAA} +30.7 & 0.005\\
\org{M.~mulatta}   & occurrences & Chr7     & 419 &  2{,}957 & 0.128 & 0.079 & \ty{AAA} $-$13.3 & 0.007\\
\org{M.~mulatta}   & occurrences & Chr14    & 349 &  2{,}090 & 0.129 & 0.087 & \ty{AAA} $-$12.7 & 0.014\\
\org{M.~mulatta}   & occurrences & Chr10    & 260 &  1{,}701 & 0.137 & 0.099 & \ty{AAA} $-$14.0 & 0.020\\
\org{Z.~mays}      & occurrences & Unplaced &  58 &     98 & 0.274 & 0.171 & \ty{ABC} +29.3 & 0.004\\
\org{A.~thaliana}  & occurrences & Chr2     & 636 &  2{,}171 & 0.068 & 0.048 & \ty{ABC} $-$7.0  & 0.011\\
\org{A.~thaliana}  & occurrences & Chr5     & 1{,}067 & 3{,}309 & 0.042 & 0.035 & \ty{ABC} +3.2 & 0.023\\
\org{A.~thaliana}  & occurrences & Chr3     & 801 &  2{,}696 & 0.048 & 0.042 & \ty{ABA} +2.5 & 0.033\\
\org{A.~thaliana}  & occurrences & Chr4     & 734 &  2{,}398 & 0.046 & 0.044 & \ty{AAB} +2.1 & 0.048\\
\bottomrule
\end{tabular}
\end{table}

\subsection{Ranking genomes by within-genome similarity}\label{sec:ranking}

Table~\ref{tab:rank} ranks the 14 genomes with testable components by the median distance of those components from
the whole proteome, for distinct trigrams (\S\ref{sec:between}). Excluding fly and human, which are described above, the three
genomes at each end are examined next. The bottom of the ranking is crowded: \org{C.~elegans} (0.0772) is only just ahead of
rat (0.0769), zebrafish (0.0764) and mouse (0.0756), so third place from the bottom should not be read as a
sharp distinction.

The raw distance depends on the size of the component. Over all 199 components it falls as the number of distinct trigrams
rises (Spearman $\rho=-0.61$), whereas its ratio to the random-gene range does not ($\rho=0.07$). The ranking therefore
says as much about how many trigrams a chromosome holds as about how its pie is composed, and the ratio column of
Table~\ref{tab:rank} is the fairer guide to whether a genome's chromosomes are unusually variable. All 14 genomes have a
median ratio below 1.

\begin{table}[t]
\centering
\caption{Genomes ranked by how closely their components follow the whole-proteome pie (distinct trigrams). Comp.: tested components.
Trigrams: median number of distinct trigrams per component. Median JS: median distance from the whole proteome. Ratio: median of
that distance divided by the 95th percentile for random gene sets. Inside: components inside the random-gene range. $q<0.05$:
components that differ after correction, for distinct trigrams (dist.) and occurrences (occ.). \org{E.~coli} K-12 and
\org{Tetrahymena} have no components to test.}
\label{tab:rank}
\small
\begin{tabular}{r l r r r r r r r l}
\toprule
& & & & Median & & & \multicolumn{2}{c}{$q<0.05$} &\\
\cmidrule(lr){8-9}
Rank & Organism & Comp. & Trigrams & JS & Ratio & Inside & dist. & occ. & Treated in\\
\midrule
1 & \org{S.~pombe} & 3 & 453 & 0.0346 & 0.58 & 3 & 0 & 0 & \S\ref{sec:most}\\
2 & \org{A.~thaliana} & 5 & 764 & 0.0374 & 0.77 & 5 & 0 & 4 & \S\ref{sec:most}\\
3 & \org{D.~melanogaster} & 6 & 728.5 & 0.0499 & 0.70 & 6 & 0 & 0 & \S\ref{sec:fly}\\
4 & \org{Z.~mays} & 11 & 461 & 0.0518 & 0.75 & 9 & 0 & 1 & \S\ref{sec:most}\\
5 & \org{X.~laevis} & 19 & 901 & 0.0606 & 0.76 & 15 & 0 & 0 & \\
6 & \org{H.~sapiens} & 25 & 573 & 0.0631 & 0.64 & 23 & 2 & 1 & \S\ref{sec:human}\\
7 & \org{M.~mulatta} & 23 & 524 & 0.0699 & 0.70 & 22 & 0 & 4 & \\
8 & \org{O.~sativa} & 12 & 308.5 & 0.0715 & 0.70 & 11 & 0 & 0 & \\
9 & \org{M.~musculus} & 22 & 537 & 0.0756 & 0.71 & 20 & 1 & 0 & \\
10 & \org{D.~rerio} & 26 & 552.5 & 0.0764 & 0.69 & 25 & 1 & 0 & \\
11 & \org{R.~norvegicus} & 23 & 404 & 0.0769 & 0.77 & 21 & 0 & 0 & \\
12 & \org{C.~elegans} & 6 & 508.5 & 0.0772 & 0.75 & 6 & 0 & 0 & \S\ref{sec:least}\\
13 & \org{E.~coli} O157:H7 & 2 & 394 & 0.0882 & 0.71 & 2 & 0 & 0 & \S\ref{sec:least}\\
14 & \org{S.~cerevisiae} & 16 & 91.5 & 0.0941 & 0.58 & 16 & 0 & 0 & \S\ref{sec:least}\\
\bottomrule
\end{tabular}
\end{table}

\FloatBarrier
\section{The three genomes closest to their own pie}\label{sec:most}

The chromosomes of \org{S.~pombe}, \org{Arabidopsis thaliana} and maize lie closest to their own whole-proteome pie, with
median distances of 0.035, 0.037 and 0.052 (Table~\ref{tab:rank}). The three do not form a taxonomic group: they are a
yeast and two flowering plants, and their whole-proteome pies differ (\ty{ABC} 75\%, 66\% and 64\%; Table~\ref{tab:pct}).

\begin{figure}[tbp]
\centering
\includegraphics[width=0.92\linewidth]{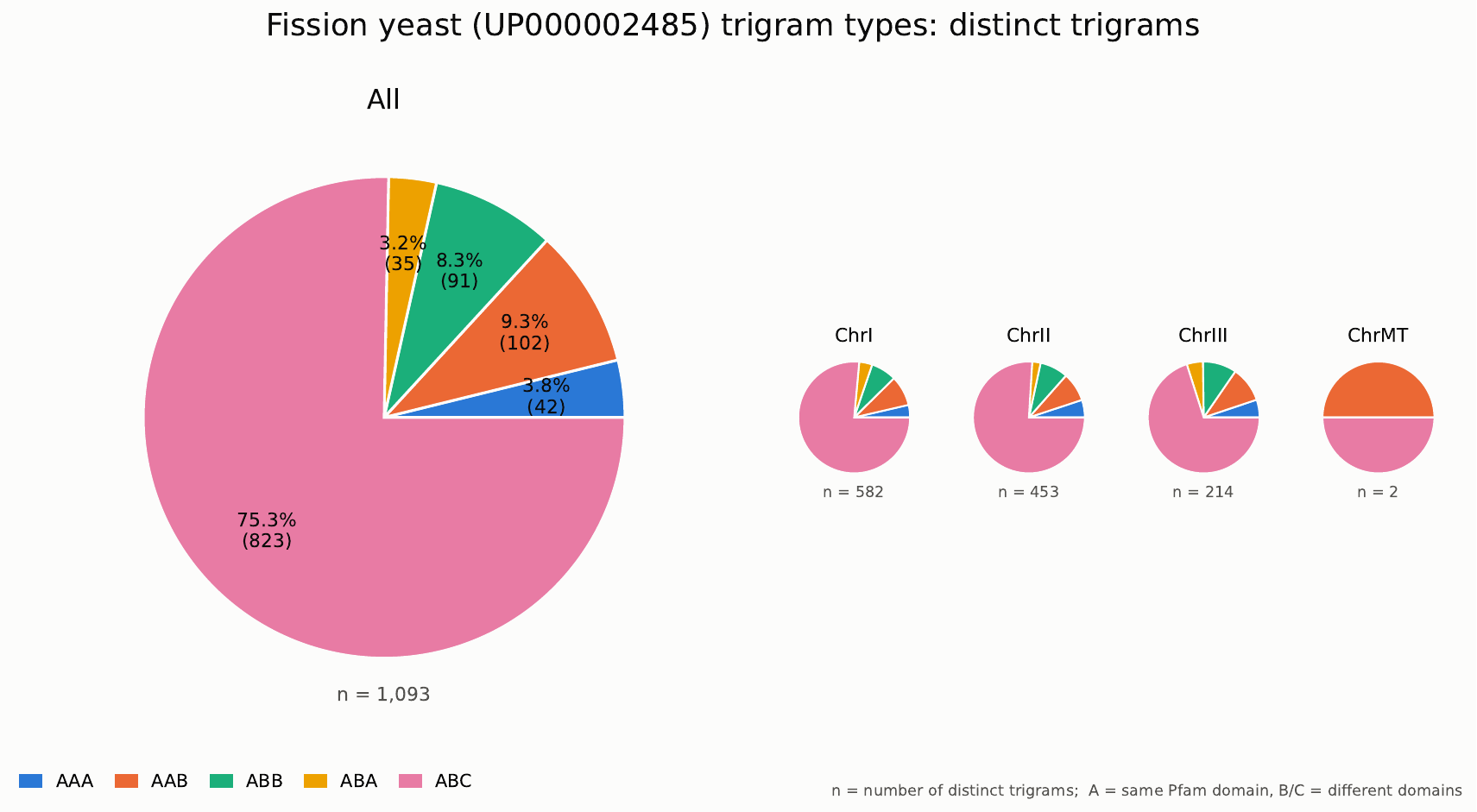}
\caption{\org{S.~pombe}: trigram types, distinct trigrams, for the whole proteome (large pie) and each component. $n$: number of
distinct trigrams. Notebook \texttt{fissionyeast\_trigram}.}
\label{fig:pombe}
\end{figure}

\paragraph{\org{Schizosaccharomyces pombe}.} The fission yeast has three chromosomes \cite{wood2002pombe}, with 124--366 genes
carrying at least three domains. Its 1{,}093 distinct trigrams are 75.3\% \ty{ABC}, 9.3\% \ty{AAB}, 8.3\% \ty{ABB}, 3.8\%
\ty{AAA} and 3.2\% \ty{ABA} (Fig.~\ref{fig:pombe}). The chromosomes are 0.022 (I), 0.035 (II) and 0.054 (III) from the whole
proteome, all inside the random-gene range; the largest difference is 5.2~pp less \ty{ABC} on chromosome~III. The occurrence
pies are also inside the range (distances 0.044--0.054, $q\ge0.64$). The largest single contributor is one threonine-rich
glycoprotein with the \org{S.~pombe}-specific repeat DUF963 (\pf{PF06131}; UniProt Q96WV6), whose 72 \ty{AAA} occurrences are
10\% of those on chromosome~II.

\begin{figure}[tbp]
\centering
\includegraphics[width=0.92\linewidth]{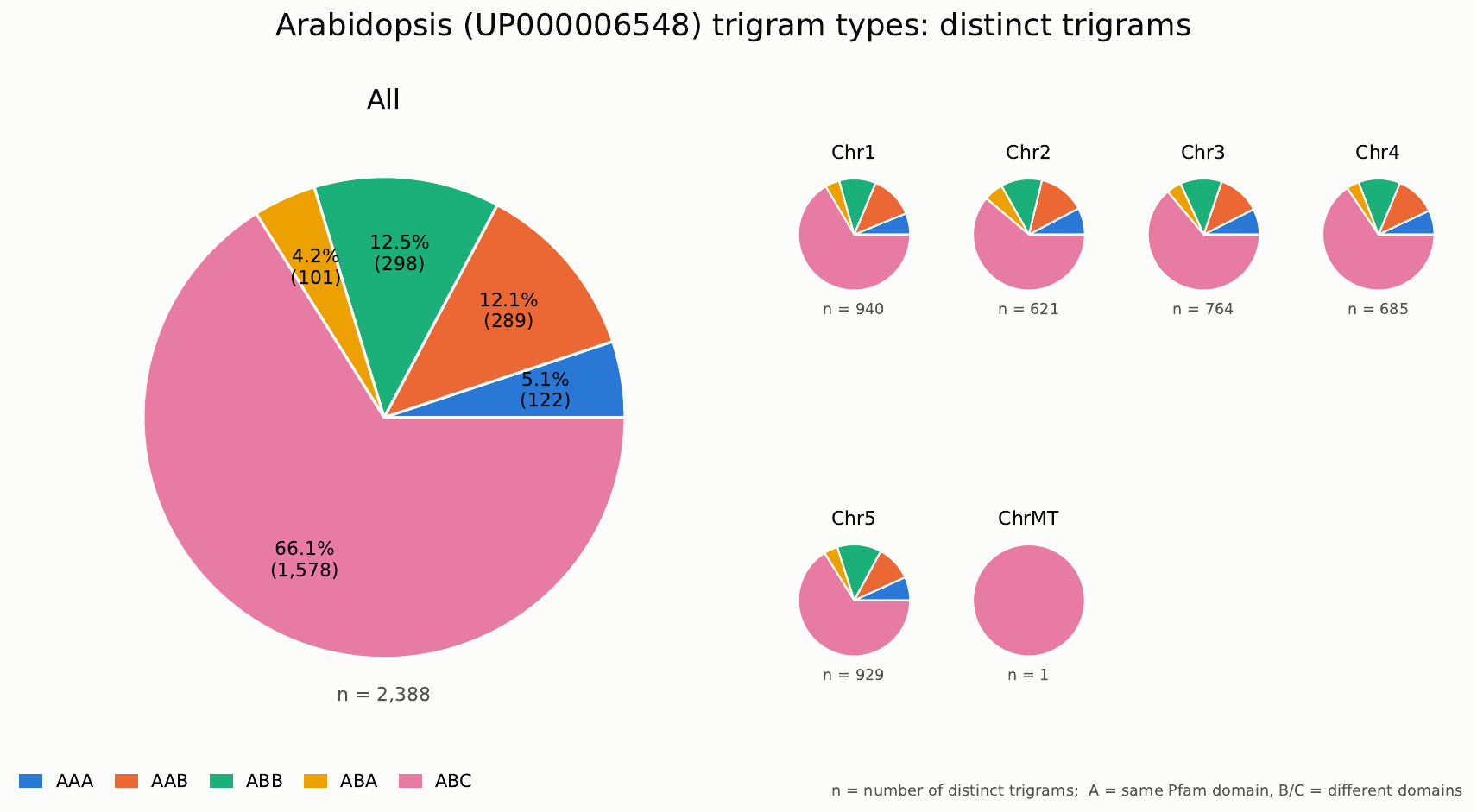}
\caption{\org{A.~thaliana}: trigram types, distinct trigrams, for the whole proteome (large pie) and each component. $n$: number of
distinct trigrams. Notebook \texttt{arabidopsis\_trigram}.}
\label{fig:arab}
\end{figure}

\paragraph{\org{Arabidopsis thaliana}.} The five chromosomes \cite{agi2000arabidopsis} give 2{,}388 distinct trigrams overall,
66.1\% \ty{ABC}, 12.5\% \ty{ABB}, 12.1\% \ty{AAB}, 5.1\% \ty{AAA} and 4.2\% \ty{ABA} (Fig.~\ref{fig:arab}). Distances run from 0.030
(chromosome~1) to 0.061 (chromosome~2), all inside the random-gene range ($q\ge0.35$). In occurrences, however, four of the five
chromosomes differ from the whole proteome (Table~\ref{tab:sig}), by distances of only 0.042--0.068. No gene dominates: the
largest single share on any chromosome is 3.7\% (\textit{WRKY19}, chromosome~4), and extensins, the repeat-rich cell-wall
glycoproteins \cite{showalter2010hrgp}, are the top contributors on chromosomes~2, 3 and 5. Because occurrences are spread
over many genes, the random-gene range is narrow (95th percentile 0.031--0.048), and small differences (at most 7~pp \ty{ABC},
on chromosome~2) become detectable. Close similarity in distinct trigrams thus coexists with statistically detectable, but
small, differences in occurrences.

\begin{figure}[tbp]
\centering
\includegraphics[width=0.92\linewidth]{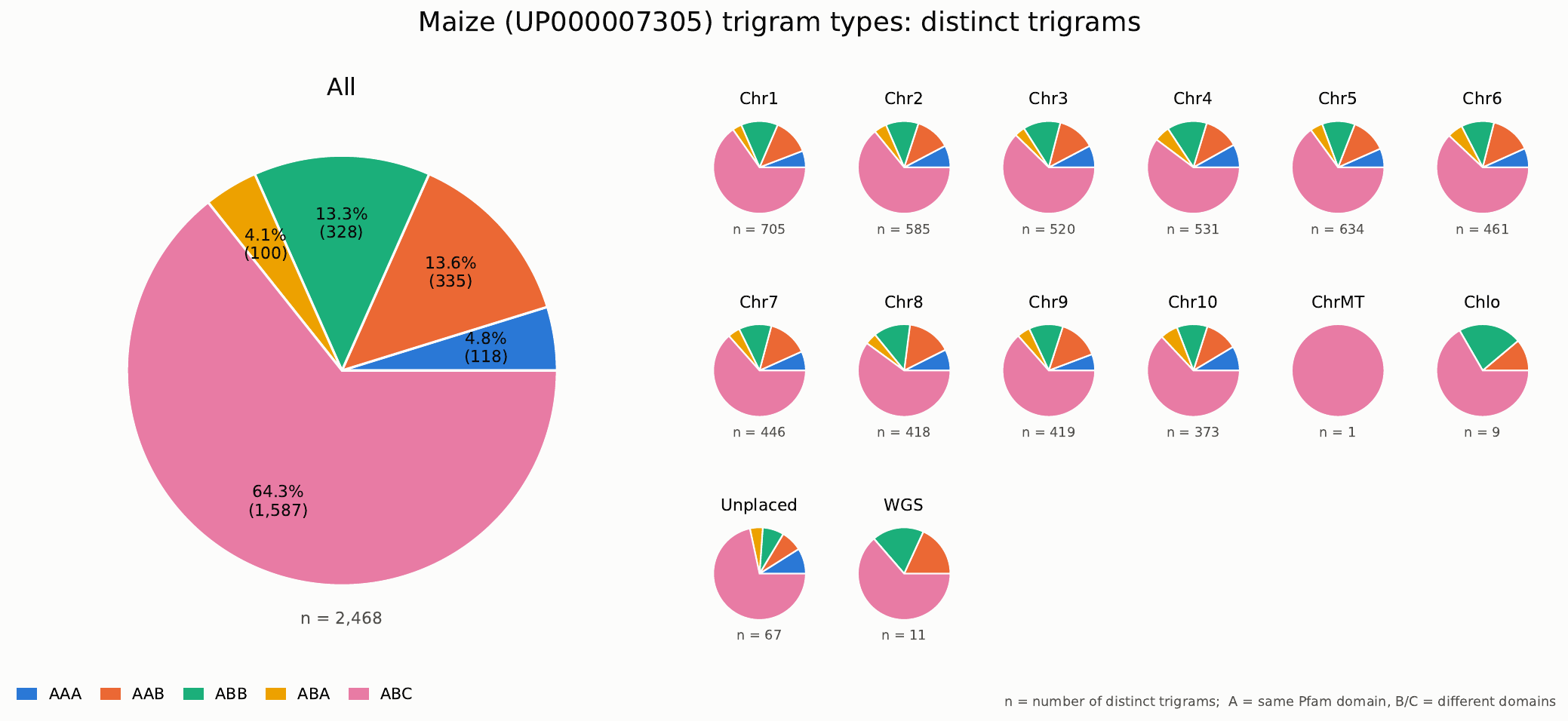}
\caption{Maize: trigram types, distinct trigrams, for the whole proteome (large pie) and each component. $n$: number of
distinct trigrams. Components with fewer than ten genes are drawn but not tested. Notebook \texttt{maize\_trigram}.}
\label{fig:maize}
\end{figure}

\paragraph{Maize.} The ten chromosomes and the unplaced proteins \cite{hufford2021maize} give 2{,}468 distinct trigrams
overall, 64.3\% \ty{ABC}, 13.6\% \ty{AAB}, 13.3\% \ty{ABB}, 4.8\% \ty{AAA} and 4.1\% \ty{ABA} (Fig.~\ref{fig:maize}). The chromosomes are
0.028--0.085 from the whole proteome, and the 58-gene unplaced set 0.137. Chromosomes~4 and 10 lie just outside the
random-gene range, but no component differs after correction ($q\ge0.25$). In occurrences only the unplaced set differs
(\ty{ABC} +29.3~pp, $q=0.004$; Table~\ref{tab:sig}); the chromosomes are 0.034--0.052 from the whole, and no single gene gives more
than 2.6\% of the occurrences on any chromosome.

\FloatBarrier
\section{The three genomes furthest from their own pie}\label{sec:least}

The chromosomes of \org{S.~cerevisiae}, \org{E.~coli} O157:H7 and \org{C.~elegans} lie furthest from their own whole-proteome
pie, with median distances of 0.094, 0.088 and 0.077 (Table~\ref{tab:rank}). ``Furthest'' is relative: every tested component of
these three genomes lies inside the random-gene range for distinct trigrams and for occurrences, and none differs
after correction. The reasons for their rank differ.

\begin{figure}[tbp]
\centering
\includegraphics[width=0.92\linewidth]{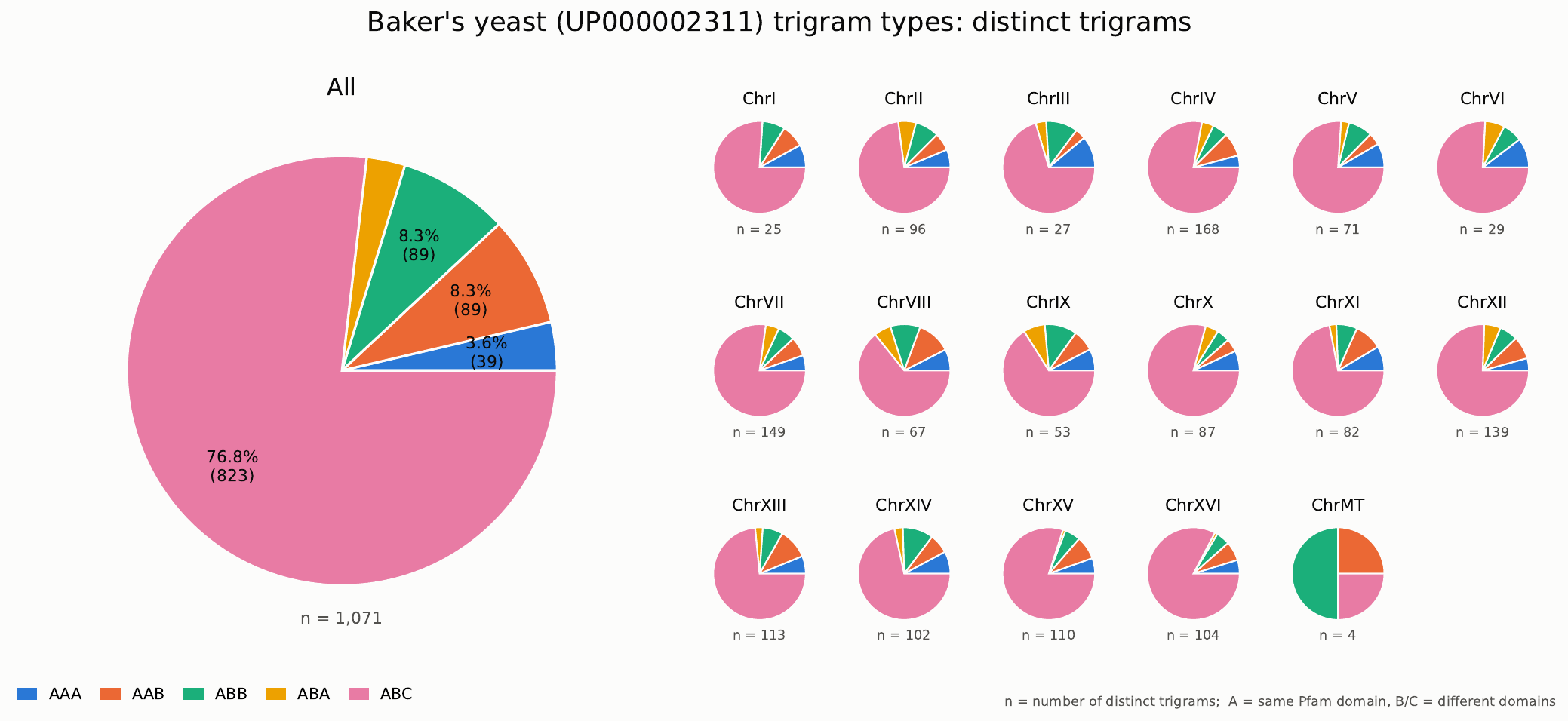}
\caption{\org{S.~cerevisiae}: trigram types, distinct trigrams, for the whole proteome (large pie) and each component. $n$: number of
distinct trigrams. ChrMT has fewer than ten genes and was not tested. Notebook \texttt{buddingyeast\_trigram}.}
\label{fig:scer}
\end{figure}

\paragraph{\org{Saccharomyces cerevisiae}: small chromosomes.} The 16 chromosomes \cite{goffeau1996yeast} carry few trigrams each
(25--168 distinct, median 91.5), so each pie is a small sample of the whole-proteome pie of 1{,}071 distinct trigrams (76.8\%
\ty{ABC}, 8.3\% each of \ty{AAB} and \ty{ABB}, 3.6\% \ty{AAA}, 2.9\% \ty{ABA}; Fig.~\ref{fig:scer}). Distances run from 0.058
(chromosome~IV) to 0.245 (chromosome~VI), and the random-gene range is correspondingly wide (95th percentile 0.11--0.30; all
$q=0.80$). Relative to that range, the budding yeast has the lowest median ratio of all genomes, equal to that of
\org{S.~pombe} (0.58). In occurrences, chromosome~I stands out (\ty{AAA} +36~pp): two flocculin genes, \textit{FLO1} and
\textit{FLO9}, whose tandem repeats vary in number \cite{verstrepen2005repeats}, give 62\% of its 52 occurrences. The difference
is not significant with 16 genes ($q=0.47$).

\begin{figure}[tbp]
\centering
\includegraphics[width=0.92\linewidth]{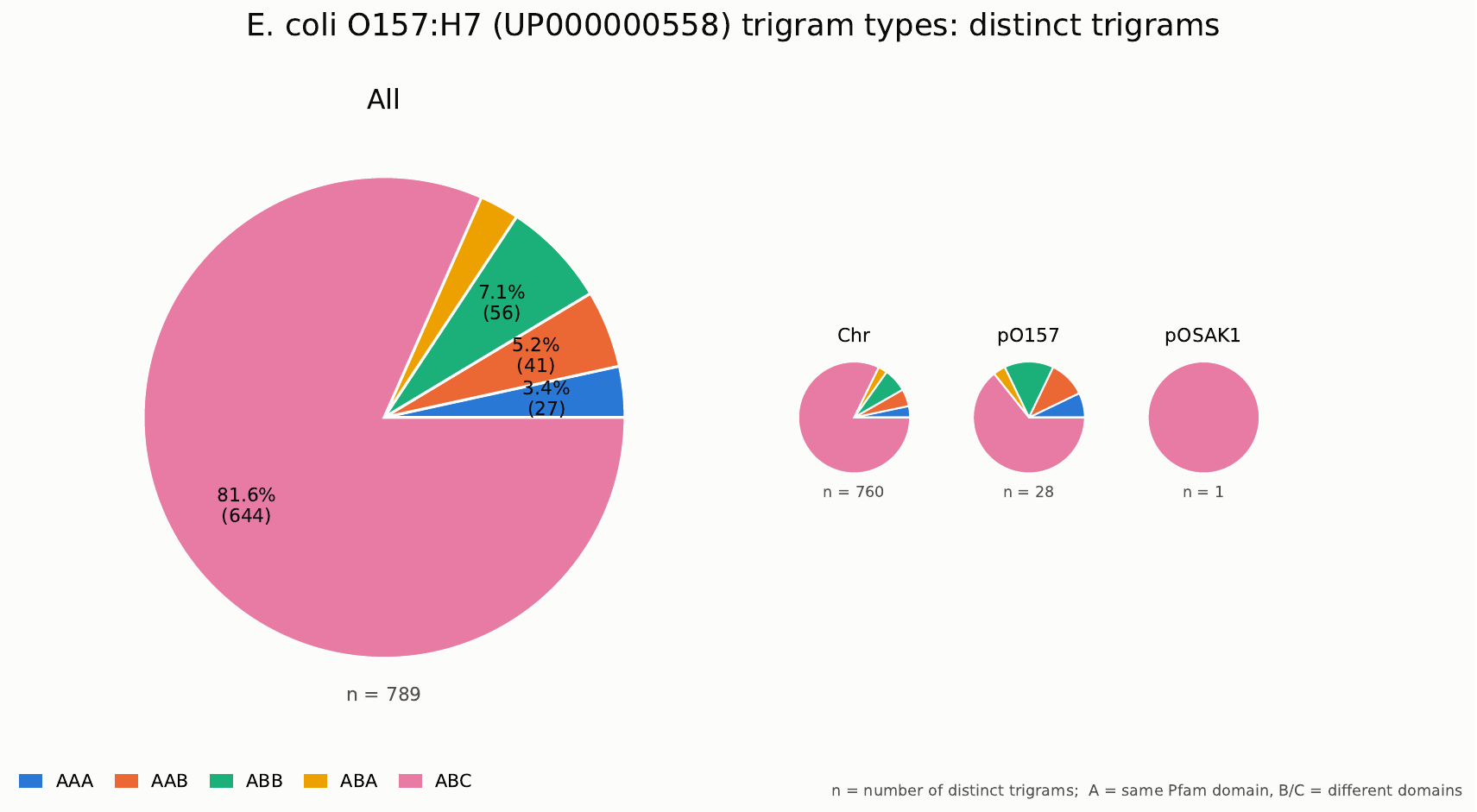}
\caption{\org{E.~coli} O157:H7: trigram types, distinct trigrams, for the whole proteome (large pie) and each component. $n$: number of
distinct trigrams. pOSAK1 has fewer than ten genes and was not tested. Notebook \texttt{ecoliO157\_trigram}.}
\label{fig:o157}
\end{figure}

\paragraph{\org{Escherichia coli} O157:H7: one small plasmid.} Only two replicons could be tested
\cite{hayashi2001o157}. The chromosome holds 760 of the 789 distinct trigrams (81.6\% \ty{ABC}, 7.1\% \ty{ABB}, 5.2\% \ty{AAB}, 3.4\%
\ty{AAA}, 2.7\% \ty{ABA}; Fig.~\ref{fig:o157}), so it essentially is the whole proteome (distance 0.007). The virulence plasmid pO157
\cite{burland1998po157} has 11 genes with 28 distinct trigrams, and is 0.169 from the whole (\ty{ABC} $-$17~pp), well inside its
random-gene range (95th percentile 0.31). The median of two values is their mean, so this genome's rank comes entirely from
the plasmid. The second plasmid, pOSAK1, has a single trigram and was not tested.

\begin{figure}[tbp]
\centering
\includegraphics[width=0.92\linewidth]{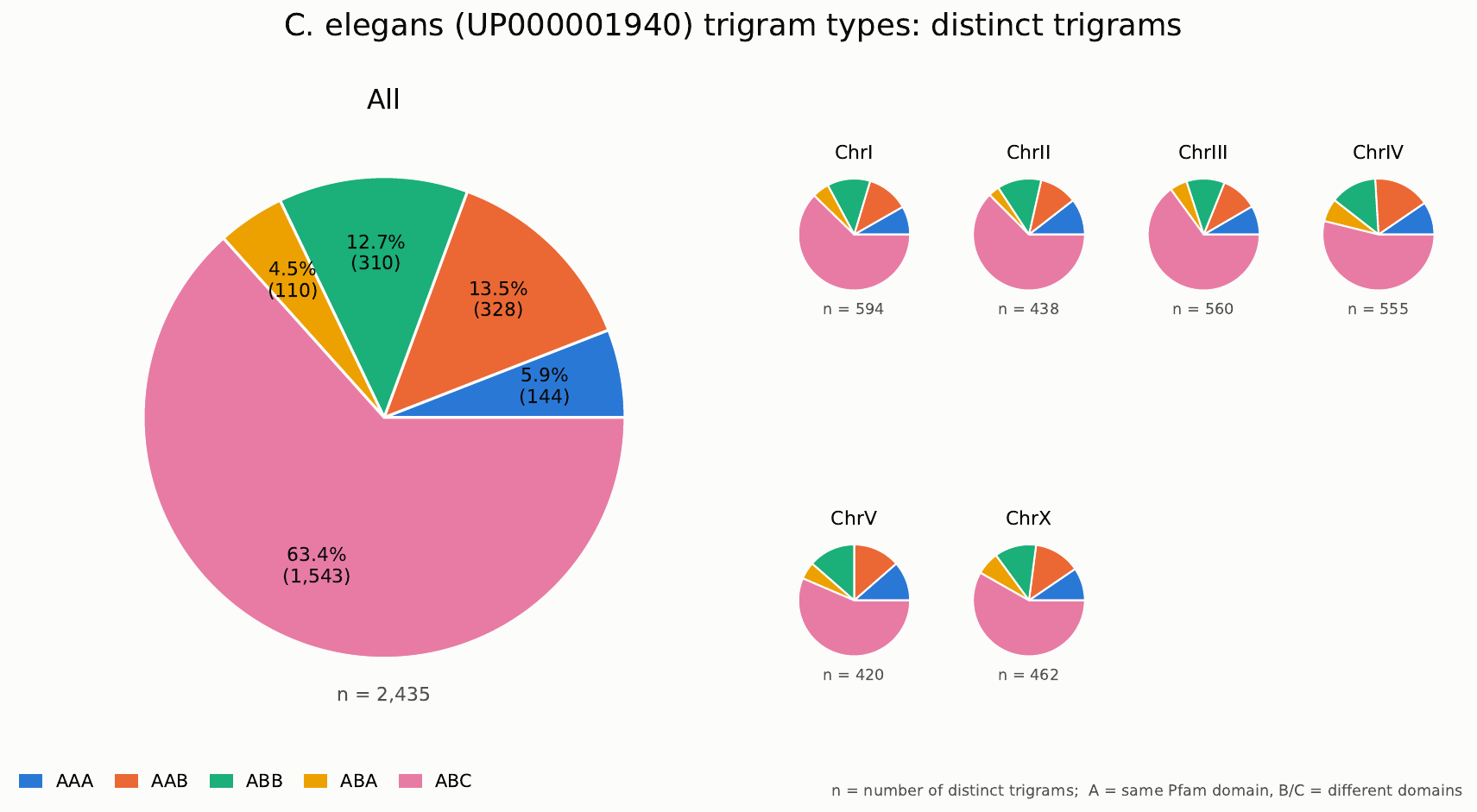}
\caption{\org{C.~elegans}: trigram types, distinct trigrams, for the whole proteome (large pie) and each component. $n$: number of
distinct trigrams. Notebook \texttt{celegans\_trigram}.}
\label{fig:cel}
\end{figure}

\paragraph{\org{Caenorhabditis elegans}: wide random-gene range.} The six chromosomes \cite{celegans1998} give 2{,}435 distinct
trigrams overall, 63.4\% \ty{ABC}, 13.5\% \ty{AAB}, 12.7\% \ty{ABB}, 5.9\% \ty{AAA} and 4.5\% \ty{ABA} (Fig.~\ref{fig:cel}). Here the
chromosomes are not small (420--594 distinct trigrams), but their distances (0.042--0.092) are high; the largest is 9.5~pp less
\ty{ABC} on chromosome~IV. The random-gene range is also wider than in \org{Arabidopsis} (95th percentile 0.097--0.107 against
0.041--0.063), so all six chromosomes remain inside it ($q\ge0.38$). In occurrences, single genes again dominate:
\textit{srap-1} gives 34\% of those on chromosome~II (624, all \ty{AAA}, across 12 isoforms), \textit{unc-89} 28\% on
chromosome~I and \textit{fln-2} 19\% on chromosome~X. None of the chromosomes differs ($q\ge0.37$).

\FloatBarrier
\section{Six years of database growth: comparison with the 2020 analysis}\label{sec:2020}

\paragraph{The comparison.} All 16 organisms were first analysed in 2020, with the original code on the UniProt proteomes and
InterProScan output of that time. Those results are kept as one sheet per organism in \texttt{data/trigram\_summary.xlsx}. Each
notebook reads its sheet and reports the Jensen--Shannon distance between the 2020 and the new pie, for the whole proteome and
per component; for the fly, the 2020 trigram files themselves were also re-categorised to compare the arms. Steps 3--6 of the
present workflow are ports of the 2020 code (\texttt{n\_gram.py}) and reproduce its output byte for byte on the 2020 input files.
The differences below therefore come from the data: the UniProt proteomes, the Pfam release, and the removal of redundant
proteins, which now keeps identical sequences once (MD5) where the 2020 rule dropped a protein whose Pfam list and length matched
those of the previous one.

\begin{figure}[t]
\centering
\includegraphics[width=\linewidth]{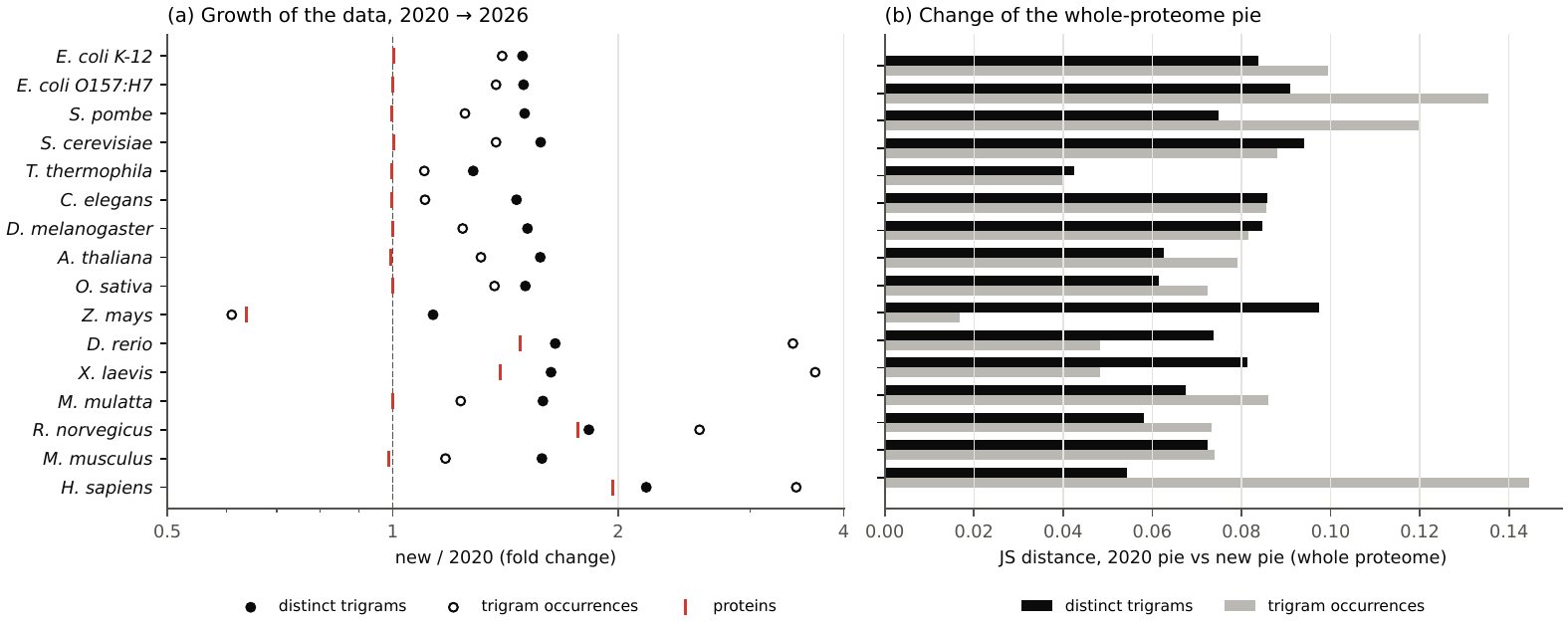}
\caption{The 2020 analysis against the 2026 re-analysis, whole proteomes of the 16 organisms. (a) Fold change of proteome entries (red ticks), distinct trigrams (filled) and trigram occurrences (open).
(b) Jensen--Shannon distance between the 2020 and the new pie. Data: \texttt{<species>\_trigram\_2020\_vs\_new.csv}; for the fly, the \texttt{fly803} sheet of \texttt{data/trigram\_summary.xlsx}.}
\label{fig:2020}
\end{figure}

\begin{figure}[t]
\centering
\includegraphics[width=\linewidth]{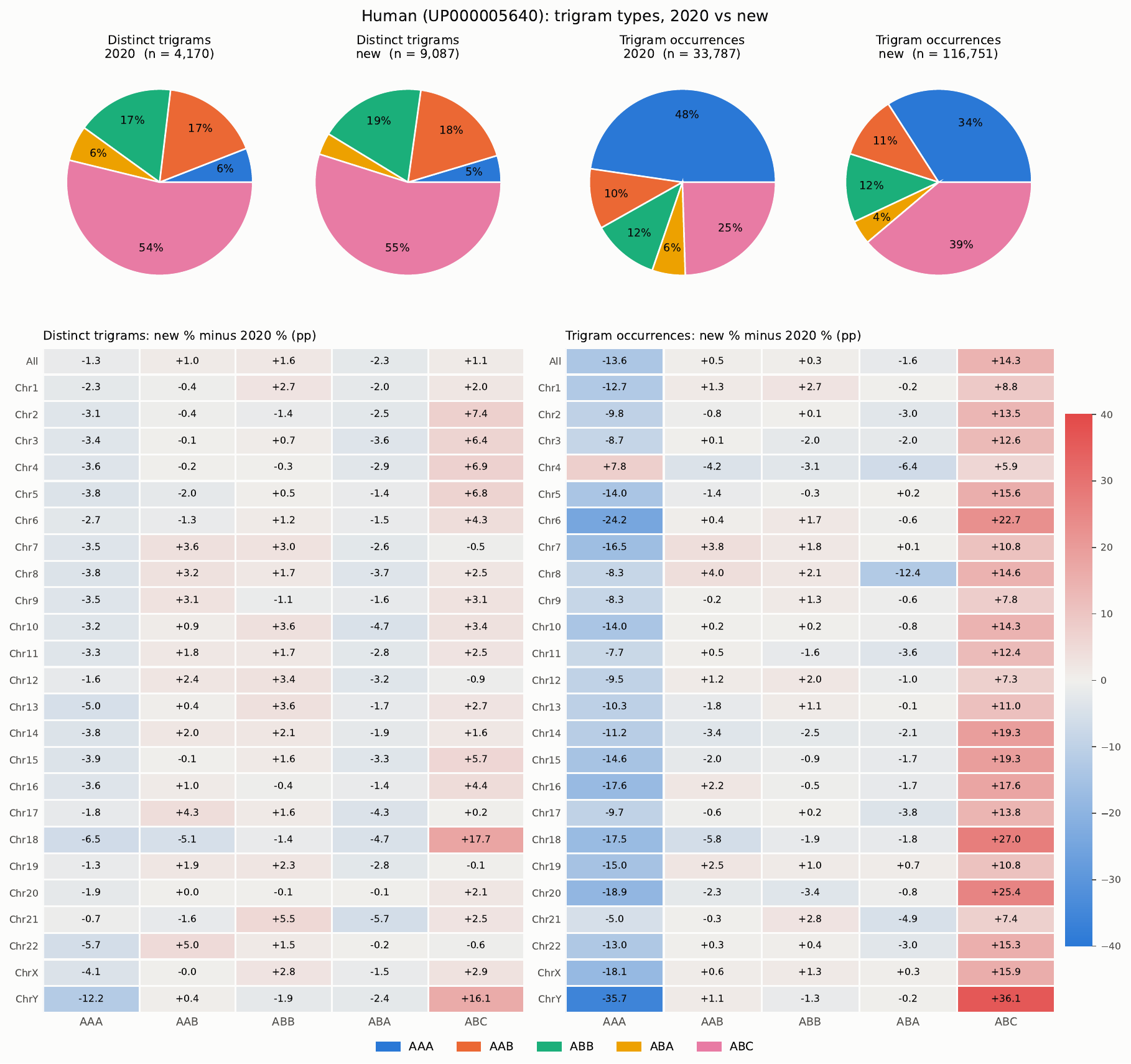}
\caption{Human, 2020 against 2026. Top: whole-proteome pies; bottom: per-chromosome change (new minus 2020, pp) for
distinct trigrams (left) and occurrences (right).}
\label{fig:human2020}
\end{figure}

\paragraph{Growth in the data.} Between 2020 and 2026 the number of Pfam families detected in each proteome rose by 20--55\%, and
the number of distinct trigrams by 13--118\% (Fig.~\ref{fig:2020}a). For the vertebrates the proteomes themselves changed: rat, zebrafish and \org{Xenopus} gained 39--76\% more
entries, mostly new isoforms, and trigram occurrences rose 2.6- to 3.7-fold, driven by isoforms of titin, nebulin and C2H2
zinc-finger genes. Maize was rebuilt on the B73 v5 assembly \cite{hufford2021maize} with 36\% fewer proteins, so its occurrences
fell by 39\%. Human combines both effects: even after excluding 116{,}508 entries not yet scanned by Pfam, the proteome doubled
(75{,}004 to 147{,}378 entries) and occurrences rose 3.5-fold (Fig.~\ref{fig:human2020}). Component sets also changed: the 2020
fly data included 23 Y-chromosome proteins that the current reference proteome lacks, and several organisms gained or lost
unplaced sets.

\paragraph{Stability of the pies.} Against this growth, the distinct-trigram pie of the whole proteome moved by a distance of only
0.042--0.098 (Fig.~\ref{fig:2020}b). Re-running the 2020 analysis on 2026 data thus changed the distinct-trigram pies by
less than they differ between multicellular eukaryotes and bacteria ($\geq$0.148). All 16 organisms gained \ty{ABC}, by 1--11~pp;
newly added Pfam families that label previously unannotated domains would produce such new combinations, although we did not test
this directly. The occurrence pie moved by 0.017--0.145, the largest change being human (\ty{AAA} 47.7\% $\to$ 34.1\%).
Six years of UniProt and Pfam releases thus changed the number of distinct trigrams by up to a factor of two and occurrences by up
to a factor of 3.7. The proportions are robust where the counts are not.

\paragraph{Sources of the differences.} Part of the 2020 \ty{AAA} excess was an artefact: the 2020 InterProScan output contained
overlapping hits of the same family in tandem repeat proteins (1.6\% of hits in fission yeast and \org{C.~elegans}; for example the
\org{S.~pombe}-specific repeat DUF963), which the current InterPro matches no longer contain. The two Pfam routes used here
(\S\ref{sec:methods}) differ by a distance of 0.013--0.023 for distinct-trigram pies on the six organisms that have both, which is
smaller than the 2020-to-2026 change for those organisms (0.04--0.09). The 2020 conclusion that distinct-trigram pies are uniform
across chromosomes is unchanged.

\FloatBarrier
\section{Discussion}

We asked whether the way proteins combine and repeat their domains, reduced to five patterns of domain trigrams, is the same
along the chromosomes of a genome. In the fruit fly, in human and in 12 further organisms, it largely is. After correction for
multiple testing, the distinct-trigram pie of a chromosome is indistinguishable from that of a random set of genes of the same
size in 98\% of the 199 cases tested, and 92\% lie inside the 95\% random-gene range. Chromosomes differ greatly in length, gene
density and evolutionary history \cite{hurst2004order}, yet the vocabulary of domain combinations they encode is drawn, in effect,
from one common distribution for each genome.

The exceptions are informative because they are few and concrete. They are places where many paralogous genes of one repeat-rich
family were made by local duplication and still sit together: the KRAB zinc-finger clusters of primate chromosome~19
\cite{grimwood2004chr19,huntley2006krab} and the zinc-finger and NLR clusters of zebrafish chromosome~4
\cite{howe2013zebrafish,howe2016nlr}. Elsewhere, single giant proteins with many isoforms (Dumpy \cite{wilkin2000dumpy}, titin
\cite{labeit1995titin}, nebulin \cite{labeit1995nebulin}, obscurin \cite{young2001obscurin}, Flo1 \cite{verstrepen2005repeats}) can
make an occurrence pie look different without making the chromosome different. Occurrence pies should therefore be tested with
genes rather than trigrams as the sampling unit, because a test on trigram counts would treat the correlated occurrences of one gene
as independent.

The comparison of genomes at the two ends of the ranking shows how the raw distance should be read. The three closest genomes are
a yeast and two plants, and they do not share a lineage. The three furthest are distant for different reasons: small chromosomes
in budding yeast, one small plasmid in \org{E.~coli} O157:H7, and a wide random-gene range in \org{C.~elegans}. In no case is the
distance outside what random gene sets give. The size of a component thus explains much of its raw distance, and a comparison
with random gene sets of the same size is needed before a chromosome can be called different.

Lying inside the random-gene range does not mean that genes are placed at random. Gene order is not random
\cite{hurst2004order}, the five-pattern pie records how much domains repeat rather than which domains or genes are present,
and the gene-level test can miss modest biases, especially on small components such as fly chromosome~4. The result is
narrower: for this one property, a chromosome behaves like a random sample of its genome's genes. Two companion analyses of
eight of the genomes studied here arrive at the same picture by other measures. The functional-element composition of most
chromosomes is a near-copy of that of the whole genome, and departs mainly where paralogues have expanded in cis, again led by
the KRAB zinc fingers of human chromosome~19 and the clusters of zebrafish chromosome~4 \cite{wu2026composition}. Members of
a gene family fall into a short-range population of tandem neighbours, whose scale follows the gene spacing of each genome,
and a long-range population consistent with random placement \cite{wu2026ideogram}. Gene clustering is therefore real but local: at the scale
of a chromosome it averages out, except where one family has expanded in place.

Across organisms the picture has two levels. Among multicellular eukaryotes (a nematode, an insect, three flowering plants and six
vertebrates) the distinct-trigram pie is closely conserved, with human the most distant member. This agrees with the view that domain
architectures follow a shared grammar whose information content is similar across eukaryotes \cite{yu2019grammar}, and extends it
to the simplest statistic of that grammar. The step to yeasts and bacteria, towards \ty{ABC}, is in the direction expected from their
fewer and shorter multidomain proteins and their fewer domain repeats \cite{ekman2005multidomain,bjorklund2006repeats}. The
near-equality of \ty{AAB} and \ty{ABB} in all 16 organisms is partly a consequence of how repeats are counted, but for distinct
trigrams it says that repeat runs are entered and left through equally diverse domains. This symmetry between the N- and C-terminal
sides of repeats would be worth testing directly, by position in the protein.

\section{Caveats}\label{sec:caveats}
\begin{itemize}[leftmargin=*]
  \item \textbf{Two Pfam sources.} Nine organisms use the Pfam~39.0 release file instead of the InterPro API. On the six
  non-fly organisms that have both, 96--98.5\% of proteins have the same domain order (\org{Tetrahymena} 91.6\%) and
  94--99.7\% of domain locations are identical; distinct-trigram pies differ by a distance of 0.013--0.023. Occurrence pies differ
  more (0.02--0.06; \ty{AAA} share $-5$ to $+6$~pp), because InterPro resolves competition between members of a Pfam clan
  \cite{finn2006clans} (WD40, EF-hand, TPR, ankyrin, LRR) differently. Occurrence shifts of a few pp in the plants and vertebrates
  are within this uncertainty.
  \item \textbf{Proteome content differs.} Human has 6.9 entries per gene and the other vertebrates 1.7--2.6, against 1.0--1.5 in the
  plants, invertebrates, yeasts and bacteria. The Pfam-date exclusion removed 76\% of mouse and 44\% of human entries. Distinct
  trigrams and gene-level tests are insensitive to isoform number; occurrence counts are not.
  \item \textbf{A trigram ignores spacing.} Domains are ordered by start position only; linker length, unannotated regions,
  overlapping domains and domains of other databases are not considered. A protein with fewer than three Pfam domains
  contributes nothing.
  \item \textbf{The test asks one question.} The gene-level permutation tests whether a chromosome's pie differs from that of
  a random gene set of the same proteome. It has little power for components with few genes (fly chromosome~4, 26 genes;
  mouse Y), and it does not compare organisms, for which only distances are reported.
  \item \textbf{The ranking of genomes is descriptive.} It uses the median raw distance, which falls with component size, depends on the
  number of tested components (two for \org{E.~coli} O157:H7) and is closely tied at the bottom (\S\ref{sec:ranking}).
  \item \textbf{Units are not uniform.} Fly arms, \org{Xenopus} homeologs and bacterial replicons are compared alongside whole
  chromosomes, organelle genomes and unplaced sets, and genes are identified by UniProt gene names.
  \item \textbf{Missing components.} The current reference proteomes contain no fly Y chromosome and no \org{Arabidopsis}
  chloroplast proteins.
\end{itemize}

\sloppy
\section*{Data and code availability}
All analyses use public releases: UniProtKB reference proteomes, release 2026\_04 \cite{uniprot2025} (proteome
identifiers in Table~\ref{tab:species}); InterPro~111.0 \cite{blum2025interpro}; and Pfam~39.0 \cite{paysanlafosse2025pfam},
including \texttt{Pfam-A.regions.tsv.gz}, \texttt{Pfam-A.hmm.dat.gz} and \texttt{interpro.xml.gz}. Data were downloaded on
5--7 November 2026. No new sequence data were generated. The workflow is documented in the notebook
\texttt{protTrigram\_workflow}; the fly analysis is in \texttt{fruitfly\_01\_download\_data} (Steps 1--2) and
\texttt{fruitfly\_02\_trigram\_analysis} (Steps 3--6, pies and similarity); and each other organism has one
notebook, \texttt{<species>\_trigram} (\texttt{ecoliK12}, \texttt{ecoliO157}, \texttt{fissionyeast},
\texttt{buddingyeast}, \texttt{tetrahymena}, \texttt{celegans}, \texttt{arabidopsis}, \texttt{rice}, \texttt{maize},
\texttt{zebrafish}, \texttt{xenopus}, \texttt{macaque}, \texttt{rat}, \texttt{mouse}, \texttt{human}), using the modules
\texttt{fruitfly\_trigram} and \texttt{species\_trigram}. Figures~\ref{fig:compsim} and \ref{fig:species} were drawn from the summary tables of these notebooks by \texttt{paper/scripts/make\_crossspecies\_figures.py}, and Figure~\ref{fig:2020} by \texttt{paper/scripts/make\_fig\_2020\_vs\_new\_v1.py}.
The source code is available from the author upon reasonable request.

\section*{Acknowledgements}
This work used no funding and no computing resources beyond a laptop. Analyses were carried out in Python with
numpy \cite{harris2020numpy}, pandas \cite{mckinney2010pandas}, scipy \cite{virtanen2020scipy}, requests and
matplotlib \cite{hunter2007matplotlib}.

The author would like to sincerely thank Anthropic's Claude Opus 5.5 for its incredibly intelligent aid in the
baking of this manuscript.

\end{document}